# A Comparative Analysis of Ensemble-Based Machine Learning Approaches with Explainable AI for Multi-Class Intrusion Detection in Drone Networks

[1]Md. Alamgir Hossain, [2]Waqas Ishtiaq, [3]Md. Samiul Islam

[1, 3]Department of Computer Science and Engineering, State University of Bangladesh, Dhaka Bangladesh

[2]Individual Researcher

[1, 3]Skill Morph Research Lab., Skill Morph, Dhaka, Bangladesh

[1]E-mail: alamgir.cse14.just@gmail.com, ORCID: https://orcid.org/0000-0001-5120-2911

[2]E-mail: waqas.ishtiaq@gmail.com, ORCID: https://orcid.org/0009-0009-8903-1308

[3]E-mail: samiulislam.cse@gmail.com, ORCID: https://orcid.org/0009-0002-0393-459X

## Abstract

The growing integration of drones into civilian, commercial, and defense sectors introduces significant cybersecurity concerns, particularly with the increased risk of network-based intrusions targeting drone communication protocols. Detecting and classifying these intrusions is inherently challenging due to the dynamic nature of drone traffic and the presence of multiple sophisticated attack vectors such as spoofing, injection, replay, and man-in-the-middle (MITM) attacks. This research aims to develop a robust and interpretable intrusion detection framework tailored for drone networks, with a focus on handling multi-class classification and model explainability. We present a comparative analysis of ensemble-based machine learning models, namely Random Forest, Extra Trees, AdaBoost, CatBoost, and XGBoost, trained on a labeled dataset comprising benign traffic and nine distinct intrusion types. Comprehensive data preprocessing was performed, including missing value imputation, scaling, and categorical encoding, followed by model training and extensive evaluation using metrics such as macro F1-score, ROC AUC, Matthews Correlation Coefficient, and Log Loss. Random Forest achieved the highest performance with a macro F1-score of 0.9998 and ROC AUC of 1.0000. To validate the superiority of the models, statistical tests, including Friedmans test, the Wilcoxon signed-rank test with Holm correction, and bootstrapped confidence intervals, were applied. Furthermore, explainable AI methods, SHAP and LIME, were integrated to interpret both global and local feature importance, enhancing model transparency and decision trustworthiness. The proposed approach not only delivers near-perfect accuracy but also ensures interpretability, making it highly suitable for real-time and safety-critical drone operations.

***Keywords***: *Drone Intrusion Detection; Unmanned Aerial Vehicles (UAV); Ensemble Learning; Explainable Artificial Intelligence (XAI); UAV Network Security; Supervised Machine Learning*

## 1. Introduction

The rapid deployment of Unmanned Aerial Vehicles (UAVs), commonly known as drones, has transformed modern applications ranging from surveillance and delivery to military and disaster management. However, the increased reliance on drone communication networks has also exposed them to a wide spectrum of cyber threats, making intrusion detection a critical challenge.

### *1.1 Background and Motivation*

In recent years, Unmanned Aerial Vehicles (UAVs), commonly referred to as drones, have evolved from niche surveillance tools to vital components of modern infrastructure across multiple sectors [1], [2]. From environmental monitoring, logistics, and smart agriculture to military reconnaissance and disaster response, UAVs are increasingly being integrated into mission-critical applications. This proliferation is largely driven by their flexibility, autonomous navigation, and ability to operate in complex or hostile environments with minimal human intervention [3], [4], [5].

However, with this advancement comes a growing dependence on secure and reliable drone communication networks. These networks often built on wireless and lightweight communication protocols are inherently susceptible to various forms of cyber intrusions, including Denial of Service (DoS), spoofing, replay, and man-in-the-middle (MITM) attacks. Such vulnerabilities can lead to catastrophic consequences, including unauthorized data access, hijacking of flight control, disruption of surveillance streams, and even physical crashes [6], [7]. Traditional cybersecurity mechanisms, while effective in general IT infrastructures, often fall short when applied to UAV systems due to real-time processing constraints, energy limitations, and the dynamic nature of drone networks [8], [9].

This growing threat landscape has necessitated the development of Intelligent Intrusion Detection Systems (IDS) specifically tailored for UAV environments. Recent progress in machine learning (ML) and ensemble-based learning offers promising avenues for detecting subtle and complex attack patterns in drone traffic. Yet, most ML-based IDS approaches lack interpretability, making them hard to trust and adopt in critical systems where decisions must be explainable to human operators [10], [11], [12]. This motivates the integration of Explainable Artificial Intelligence (XAI) techniques, which can illuminate the internal decision logic of complex models and thereby enhance system transparency, trust, and operational safety.

### *1.2 Problem Statement*

The increasing reliance on Unmanned Aerial Vehicles (UAVs) for critical operations has exposed drone communication networks to a variety of cyber-attacks, including DoS, spoofing, and MITM attacks. Traditional security systems are ineffective in addressing the unique challenges of UAV networks, such as limited resources and dynamic topologies. Furthermore, existing intrusion detection systems often lack explainability, making it difficult for human operators to trust the model’s decisions. This research addresses the need for a high-performance, multi-class IDS that can detect a wide range of attacks in real-time while providing explainable insights to ensure trust and transparency in drone operations.

### *1.3 Research Objectives*

The primary objective of this research is to develop an intelligent and explainable intrusion detection system (IDS) for drone communication networks. Specifically, the research aims to:

- Design and implement a multi-class IDS using ensemble-based machine learning models, such as Random Forest, XGBoost, and CatBoost, to detect various types of cyber-attacks on drone networks.
- Enhance model interpretability by integrating Explainable AI (XAI) techniques, such as SHAP and LIME, to ensure transparency and trust in the model’s decision-making process.
- Evaluate the performance of the IDS through extensive cross-validation and statistical analysis, focusing on accuracy, precision, recall, and F1-score across different attack types.
- Conduct an ablation study to identify the most influential features and assess the impact of feature selection on the overall performance of the IDS.

These objectives aim to provide both high accuracy and model transparency, making the IDS suitable for real-time deployment in secure drone networks.

### *1.4 Contributions*

This research makes several key contributions to the field of intrusion detection for drone networks:

- Development of a Multi-Class IDS: We propose a robust ensemble-based machine learning framework for detecting various types of cyber-attacks targeting drone communication systems. The framework is capable of handling multiple attack classes and ensuring high detection accuracy across different attack scenarios.
- Integration of Explainable AI (XAI) Techniques: By incorporating SHAP and LIME, we provide interpretability and transparency to the IDS, enabling human operators to understand the reasoning behind the model's decisions. This is crucial for trust and accountability in operational drone systems.

- Comprehensive Performance Evaluation: Through rigorous cross-validation and statistical testing, we assess the performance of multiple ensemble classifiers and provide comparative analysis to identify the most effective model for intrusion detection in drone networks.
- Feature Importance Analysis: An ablation study is conducted to highlight the key features that drive the model's decision-making process, offering insights into the most critical parameters for detecting different types of attacks.

### *1.5 Organization of the Paper*

The paper is structured as follows: Section 2 reviews related work on intrusion detection systems for drone networks. Section 3 details the methodology, including the dataset, preprocessing, model selection, and explainable AI integration. Section 4 presents the results and discussion, evaluating the performance of the proposed models. Finally, Section 5 concludes the paper with a summary of findings and suggestions for future research.

## 2. Related Works

The rapid adoption of Unmanned Aerial Vehicles (UAVs) across industries has necessitated the development of security frameworks to protect the integrity and confidentiality of drone communication networks. However, the dynamic and resource-constrained nature of these networks makes it challenging to apply traditional security measures. This section explores the existing literature on intrusion detection systems (IDS) for UAV networks, highlighting the methods, techniques, and challenges identified in recent studies.

### *2.1 Intrusion Detection in UAV Networks*

UAVs are typically integrated into complex, wireless networks that are highly susceptible to cyber-attacks. Several research studies have investigated the use of IDS specifically tailored for UAV environments. Traditional IDS methods, often based on signature detection, fail to address the unique characteristics of drone communication, such as mobility, limited processing power, and dynamic network topologies [13], [14]. Consequently, anomaly-based detection methods have become the preferred approach in the context of UAV networks. These methods monitor network traffic in real-time, identifying any deviations from established patterns that could indicate malicious activity.

UAV network-based intrusion detection, Abdullah and Mishra [15] proposed a machine learning-based framework for detecting Denial of Service (DoS) attacks in drone communication. They employed k-Nearest Neighbors (k-NN) with an accuracy of 55%, Decision Tree (DT) 97%, Random Forest (RF) 97% to classify normal and malicious traffic. However, their approaches did not address the multi-class nature of drone attacks and lacked scalability for handling large datasets. Aldaej et al. [16] present a framework focused on improving the security of drone-based networks. The authors explore the integration of IoT with drones and utilize machine learning (ML) techniques, specifically hybrid logistic regression and random forest, to detect cybersecurity threats. The proposed system aims to secure drone networks by analyzing data from IoT sensors, applying machine learning models for attack detection, and ensuring data privacy and integrity through blockchain technology. The experimental results show that the proposed model achieves significant performance, including high accuracy (98.58%) and recall (98.59%). The paper includes a focus on only four attack categories (DoS, r2l, u2r, and probe attacks) and its scalability challenges when applied to a larger number of drones or data instances.

### *2.2 Ensemble Learning in IDS for UAV Networks*

Ensemble methods, which combine multiple learning algorithms to improve prediction accuracy, have been gaining traction in IDS research for UAV networks. These methods, including Random Forest, Extra Trees, and XGBoost, have shown superior performance in handling the complexity and variety of attacks in network security. They excel at classifying data with high-dimensional feature spaces and are robust to overfitting, making them ideal candidates for real-time, large-scale IDS applications [17].

Lin et al. [18] introduce an innovative approach to address the issue of imbalanced datasets in UAV intrusion detection. By combining Improved Stratified Sampling (ISSEL) with ensemble learning, the method enhances the feature space representation and balances normal and attack data. The approach, using classifiers like Decision Trees, Extra Trees,

Random Forest, GBDT, and XGBoost, achieves high detection accuracy (99.42%) and strong recall (98.28%), outperforming traditional methods like SMOTE and ADASYN. However, while recall and F1 score were excellent, precision could be further improved, and the method's computational cost was higher compared to lightweight models.

*2.3 Explainable AI for IDS in UAV Networks*

One of the significant challenges in deploying machine learning-based IDS is the lack of transparency in model decision-making, which is particularly critical in security-sensitive domains like UAV networks. To address this issue, Explainable Artificial Intelligence (XAI) techniques, such as SHAP (SHapley Additive exPlanations) and LIME (Local Interpretable Model-agnostic Explanations), have been incorporated into IDS frameworks [19]. These techniques allow practitioners to understand which features contribute most to the model's decisions, enhancing the trustworthiness of the system. Researchers use [20], [21] LIME was used to explain the predictions of black-box models in cybersecurity applications, including intrusion detection. Their results demonstrated that LIME could effectively provide interpretable explanations for complex model predictions, making them more actionable for security analysts. Similarly, SHAP values have been used to provide global and local feature importance scores, shedding light on the internal workings of ensemble classifiers in UAV network security.

*2.4 Limitations of Current Approaches*

Despite advancements in IDS for UAVs, several limitations remain. Scalability remains a major concern, as traditional machine learning models often struggle with the sheer volume of traffic data generated by drones, especially in real-time applications. Additionally, multi-class detection, identifying multiple types of attacks within the same network session remains a challenge. Most studies have focused on binary classification, which fails to address the complexity of real-world attacks where multiple threats may be active simultaneously.

Furthermore, many existing models lack robustness to evolving attack strategies. Cybercriminals continuously develop new techniques to evade detection, and IDS models need to adapt quickly to emerging threats. Another limitation is the computational cost of certain machine learning techniques, which can be prohibitive in resource-constrained UAV environments.

*2.5 Gaps in Existing Research*

While previous research has made significant strides in UAV IDS, there are still notable gaps. Firstly, the use of multi-class classification to detect a broader range of attacks in real-time drone networks is underexplored. Additionally, integrating explainable AI into these models has not been sufficiently addressed, which limits their adoption in critical, real-world applications where operators need clear insights into model decisions. Moreover, real-time deployment in resource-constrained environments, where latency and power efficiency are paramount, requires further optimization of both the detection algorithms and the model architectures.

This research addresses the aforementioned gaps by proposing a multi-class intrusion detection system for drone networks, using ensemble machine learning models combined with explainable AI techniques. The proposed approach offers several advantages over existing methods, including high performance in detecting a broad range of cyber-attacks, interpretability for security analysts, and a scalable framework suitable for real-time deployment in resource-constrained UAV environments. Our research also emphasizes the importance of feature selection and engineering to optimize model accuracy, providing insights into the most critical parameters for identifying drone network threats.

## 3. Methodology

This section outlines the complete pipeline used for developing and evaluating the proposed multi-class drone intrusion detection system. It includes data preprocessing, model training, performance evaluation, statistical analysis, and explainable AI integration.

*3.1 Dataset*

This research utilized the ISOT Drone Anomaly Detection Dataset [22], a rich and diverse multi-class dataset specifically designed for intrusion detection in drone communication networks. The dataset comprises over 14 hours of malicious traffic and 10 hours of benign traffic, originally captured in PCAP format and subsequently converted into structured CSV files to facilitate machine learning applications. To prepare the data, a structured parsing routine was implemented using Python's os and pandas libraries. Each subdirectory in the dataset corresponds to a specific class label either a benign scenario or a distinct type of attack. CSV files from each directory were read, assigned a class label based on their folder name, and then aggregated into a unified DataFrame. This enabled the creation of a consistent and labeled dataset containing both benign and malicious drone communication samples.

The final dataset encompassing 10 traffic categories, including one benign class and nine distinct attack classes. These classes represent realistic cyber threats observed in UAV networks and reflect various stages of the attack surface in drone communication protocols. The Benign class includes normal drone traffic without any malicious behavior. The DoS attacks category comprises multiple subtypes such as UDP Flood, TCP Flood, HULK, Slowloris, and SlowHTTPTest which aim to overload the network or system resources of the drone. Deauthentication attacks manipulate Wi-Fi management frames to disconnect the drone from its legitimate controller. MITM (Man-In-The-Middle) attacks exploit vulnerabilities by intercepting, modifying, or injecting packets between the drone and the command system.

Further, Injection attacks introduce malicious payloads into the communication stream, while IP Spoofing disguises malicious nodes using forged IP addresses. Replay attacks involve retransmitting legitimate packets to simulate authorized behavior and cause confusion. Video Interception targets the visual surveillance capabilities of the drone by hijacking its video transmission feed. Payload Manipulation changes the data segment within packets to alter drone instructions, and Unauthorized UDP Packets refer to sending illegitimate or unsolicited UDP traffic to disrupt normal communication.

Each instance in the dataset includes a variety of raw and derived features, such as Payload Length, Entropy, Inter-Arrival Time, and Drone_port, which are crucial for modeling traffic behavior. The inclusion of these detailed and heterogeneous attack types makes the dataset ideal for training and evaluating machine learning models on multi-class classification tasks. Moreover, the labeled structure and diversity of attacks ensure the development of robust, explainable, and generalizable intrusion detection systems tailored for drone environments.

*3.2 Data Preprocessing*

The performance and reliability of any machine learning-based intrusion detection system heavily depend on rigorous data preprocessing. In this study, comprehensive data cleaning, normalization, and encoding procedures were applied to prepare the dataset for multi-class classification. The steps are outlined below:

*3.2.1 Missing Value Handling*

To ensure model robustness and prevent bias due to incomplete data, we systematically identified and treated missing values. Let the dataset be denoted as $D = \{(x)^i, (y)^i\}_{i=1}^{N}$, where, $(x)^i \epsilon R^d$ represents the feature vector and $(y)^i \epsilon \{1, \dots, K\}$ the corresponding class label. Missing entries in the feature matrix $X \in R^{N\times d}$ were processed as follows:

i) Numerical Features: Missing values were imputed using the median of each respective feature: $x_j^{(i)} \leftarrow median\left(\{x_j^{(k)} : x_j^{(k)} \neq NaN\}\right)$ for all $x_j^{(i)} = NaN, j \in N$. Where, $N$ denotes the set of numerical feature indices.

ii) Categorical Features: Missing values were filled using the mode (most frequent value) for each column: $x_j^{(i)} \leftarrow mode\left(\{x_j^{(k)} : x_j^{(k)} \neq NaN\}\right)$ for all $x_j^{(i)} = NaN$, $j \in C$. Where, $C$ is the set of categorical feature indices. After imputation, the dataset was validated to confirm that $\sum_{i=1}^{N} \sum_{j=1}^{d} 1\left\{x_j^{(i)} = NaN\right\} = 0$.

*3.2.2 Feature Scaling*

For numerical stability and convergence during model training, all continuous features were standardized using Z-score normalization, computed as: $z_j^{(i)} = \frac{x_j^{(i)} - \mu_j}{\sigma_j}$. Where, $\mu_j$ and $\sigma_j$ denote the mean and standard deviation of feature $x_j$ . This transformation ensures each feature has a mean of 0 and unit variance, improving the performance of distance-sensitive models.

*3.2.3 Categorical Encoding*

Categorical features (e.g., Protocol Type, DS status) were transformed into numerical format using label encoding. Each category in a column $x_j \in C$ was mapped to a unique integer: $Encode(v) = k, v \in Domain(x_j)$. This mapping preserved the class separability required for tree-based models without introducing the sparsity of one-hot encoding.

To ensure consistent encoding, all categorical values were explicitly cast to string type before applying *LabelEncoder*.

*3.2.4 Label Construction*

The Label column was derived directly from the folder name during data ingestion, representing the class $y^{(i)} \in Y = \{0,1, \ldots ,9\}$ with each integer corresponding to one of the ten attack types or benign traffic. This column served as the target variable throughout training and evaluation.

*3.2.5 Final Feature Matrix*

After preprocessing, the final dataset consisted of: N = 301555 samples, d = 63 preprocessed features, No missing values, All features numerically encoded and standardized. This processed matrix $X \in R^{N \times d}$ and label vector $y \in \{0,1, \ldots ,9\}^N$ were then used for model training and evaluation.

After preprocessing, the final dataset revealed a significant class imbalance, with benign and certain attack types dominating the distribution shown in ***Figure 1***. This imbalance was preserved to reflect real-world UAV traffic scenarios and was handled during model training using class-weighted strategies. Each numeric label in the dataset corresponds to a specific traffic class. The mapping is as follows:

0 = Benign, 1 = DoS Attacks, 2 = Injection, 3 = IP Spoofing, 4 = MITM, 5 = Password Cracking, 6 = Payload Manipulation, 7 = Replay Attack, 8 = Unauthorized UDP Packets, 9 = Video Interception Attack.

This mapping was preserved during preprocessing and used consistently across training, evaluation, and explainability analysis.

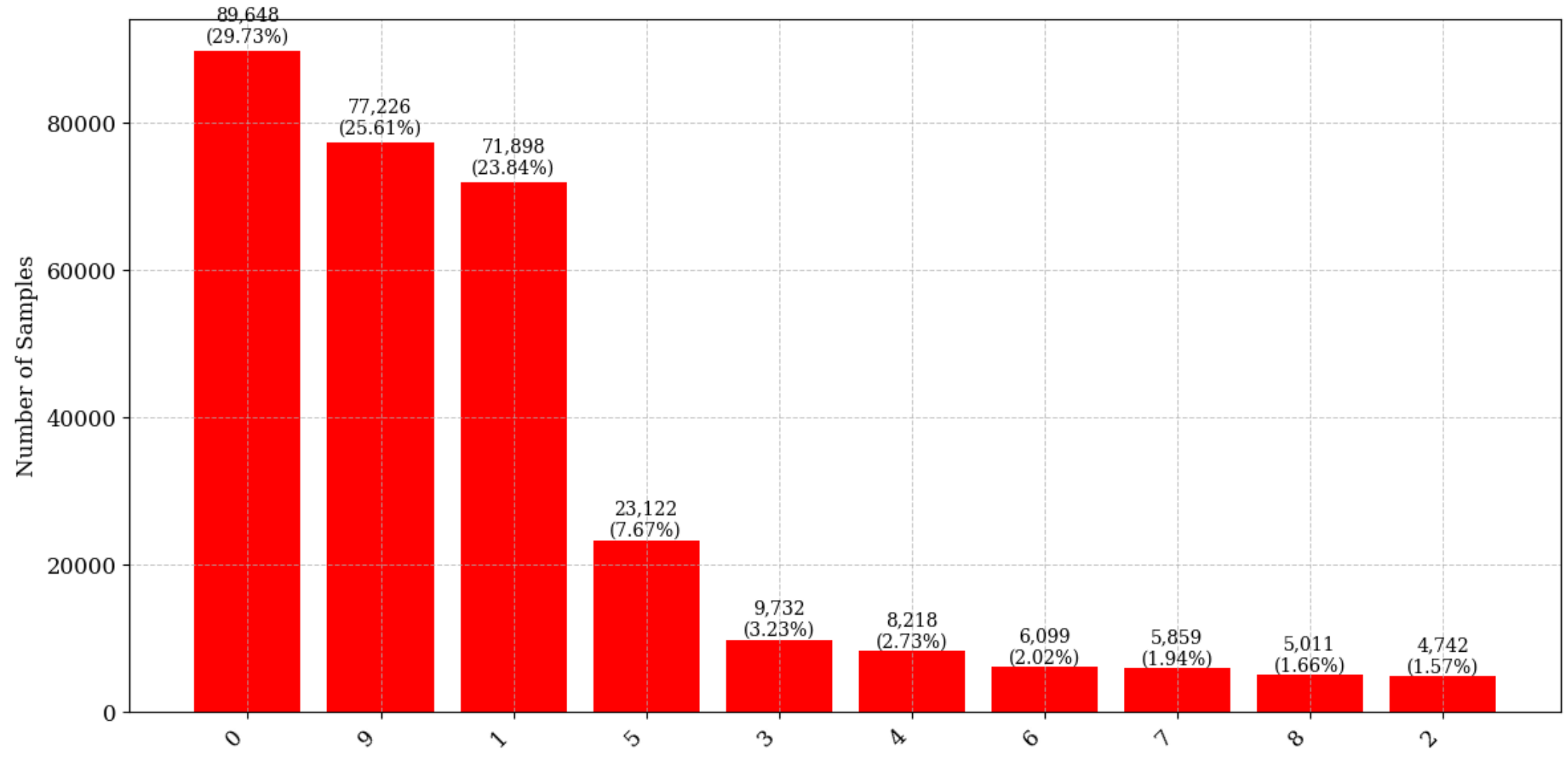

**Figure 1** Class-Wise Distribution of Samples after Data Preprocessing

*3.3 Ensemble Model Selection and Training*

To perform multi-class intrusion detection in drone networks, we employed five state-of-the-art ensemble-based classifiers: Random Forest, Extra Trees, AdaBoost, CatBoost, and XGBoost. These models were selected due to their high classification performance, ability to handle complex feature interactions, and compatibility with explainable AI methods. All models were trained using the preprocessed dataset.

*3.3.1 Random Forest Classifier*

Random Forest (RF) is an ensemble of decision trees trained on bootstrapped samples of the data with feature randomness introduced at each split. The prediction is based on majority voting over $T$ trees:

$f_{RF}(x) = \arg\max_{y \epsilon Y} \sum_{t=1}^{T} 1\{h_t(x) = y\}$, Where, $h_t$ is the prediction of the $t^{th}$ decision tree.

***Algorithm 1**: Random Forest Classifier*

1. For $t = 1$ to $T$:
   a) Sample with replacement from the training set.
   b) Build a decision tree using a random subset of features.
2. Aggregate predictions from all trees by majority vote.

*3.3.2 Extra Trees Classifier*

Extra Trees (Extremely Randomized Trees) differs from RF by using the entire training set (no bootstrap sampling) and introducing more randomness by choosing split thresholds randomly rather than by optimizing for the best split.

Prediction is similarly based on: $f_{ET}(x) = \arg\max_{y \epsilon Y} \sum_{t=1}^{T} 1\{h_t(x) = y\}$.

***Algorithm 2**: Extra Trees Classifier*

1. For each tree $t$:
   a) Use the entire dataset (no bootstrapping).
   b) Select a random feature and a random split value.
   c) Build the tree using these random thresholds.
2. Predict using majority vote.

*3.3.3 AdaBoost Classifier*

AdaBoost (Adaptive Boosting) combines weak learners $h_t$ sequentially, adjusting sample weights to focus on previously misclassified instances. The final model is a weighted majority vote:

$f_{AB}(x) = argy \in \max_{y \epsilon Y} \sum_{t=1}^{T} \alpha_t \cdot 1\{h_t(x) = y\}$. Where, $\alpha_t$ is the weight assigned to the $t^{th}$ weak learner, typically calculated as: $\alpha_t = \frac{1}{2} \ln\left(\frac{1 - \varepsilon_t}{\varepsilon_t}\right)$, and $\varepsilon_t$ is the classification error of $h_t$.

***Algorithm 3**: AdaBoost Classifier*

1. Initialize equal weights for all samples.
2. For $t = 1$ to $T$:
   a) Train weak learner $h_t$ on weighted samples.
   b) Compute error $\varepsilon_t$.
   c) Compute weight $\alpha_t$.
   d) Update sample weights.
3. Predict using the weighted vote.

*3.3.4 CatBoost Classifier*

CatBoost is a gradient boosting algorithm designed to handle categorical data efficiently through ordered boosting and target statistics. It minimizes the multiclass loss function using gradient descent over decision trees:

$L(f) = \sum_{i=1}^{N} \ell\left(f(x^{(i)}, y^{(i)}\right), f(x) = \sum_{t=}^{T} \eta_t h_t(x)$. Where $\ell$ is a multi-class log loss, $\eta$ is the learning rate, and $h_t$ is the decision tree at iteration $t$.

*Algorithm 4: CatBoost Classifier*

1. Transform categorical features into numeric statistics (ordered target encoding).

2. Initialize model with base prediction.

3. For $t = 1$ to $T$:

   a) Compute gradients of loss.

   b) Fit a tree to the negative gradients.

   c) Update model: $f_t(x) = f_{t-1}(x) + \eta_t h_t(x)$

*3.3.5 XGBoost Classifier*

XGBoost implements gradient boosting with advanced regularization and parallel computation. It minimizes the regularized objective: $L(f) = \sum_{i=1}^{N} \ell\left(\hat{y}^{(i)}, y^{(i)}\right) + \sum_{t=1}^{T} \Omega(h_t)$. Where, the regularization term $\Omega(h) = \gamma T + \frac{1}{2}\lambda \sum_j w_j^2$ penalizes model complexity.

*Algorithm 5: XGBoost Classifier*

1. Initialize base prediction $f_0(x)$
2. For $t = 1$ to $T$:
   a) Compute pseudo-residuals (gradients).
   b) Fit tree $h_t(x)$ to gradients.
   c) Update model: $f_t(x) = f_{t-1}(x) + \eta_t h_t(x)$

All models were trained using 80% of the dataset, stratified by class labels to preserve class distribution. The remaining 20% was used for evaluation. Class weights were automatically computed and applied to handle class frequency variation. Default hyperparameters were used unless otherwise specified. The number of estimators was set to 100 across all models for fair comparison.

*3. 4 Evaluation Metrics*

To evaluate the performance of the proposed multi-class intrusion detection models for drone networks, we adopted a comprehensive set of evaluation metrics. These metrics not only assess classification accuracy but also provide insights into class-wise performance, model robustness, and probability calibration. Let $y = \{y^{(i)}\}_{i=1}^{N}$ be the true class labels and $\hat{y} = \{\hat{y}^{(i)}\}_{i=1}^{N}$ the predicted labels.

*3.4.1 Accuracy*

Accuracy measures the proportion of correctly classified instances over the total number of instances:

$Accuracy = \frac{1}{N}\sum_{i=1}^{N} 1\{y^{(i)} = \hat{y}^{(i)}\}$. While widely used, accuracy can be misleading in datasets with uneven class distributions. Hence, we complemented it with additional metrics.

*3.4.2 Precision (Macro-Averaged)*

Precision assesses the correctness of positive predictions. Macro-averaging computes precision independently for each class and takes the mean:

$Precision_{macro} = \frac{1}{K}\sum_{k=1}^{K}\frac{TP_k}{TP_k+FP_k}$. Where, $TP_k$ and $FP_k$ are true positives and false positives for class $k$.

*3.4.3 Recall (Macro-Averaged)*

Recall (also called sensitivity or true positive rate) evaluates the model's ability to identify all instances of a class:

$Recall_{macro} = \frac{1}{K}\sum_{k=1}^{K}\frac{TP_k}{TP_k+FN_k}$. Where, $FN_k$ is the number of false negatives for class $k$.

*3.4.4 F1-Score (Macro-Averaged)*

The F1-score is the harmonic mean of precision and recall. For macro-averaging across $K$ classes:

$F1_{macro} = \frac{1}{K}\sum_{k=1}^{K}\frac{2\cdot Precision_k\cdot Recall_k}{Precision_k+Recall_k}$. This metric is particularly useful when both false positives and false negatives carries high cost.

*3.4.5 Balanced Accuracy*

Balanced accuracy is the average of recall obtained on each class, providing a more truthful measure for datasets with disproportionate class sizes which is equivalent to macro-recall.

*3.4.6 Matthews Correlation Coefficient (MCC)*

MCC is a correlation coefficient between the observed and predicted classifications. It returns a value between −1 and +1, where +1 indicates perfect prediction and 0 means random guessing. For multiclass problems:

$MCC = \frac{\mathrm{c}\cdot\mathrm{s}-\sum_k p_k t_K}{\sqrt{(S^2-\sum_K P_K^2)(S^2-\sum_K t_K^2)}}$. Where, $c = \sum_k TP_k, p_k$ and $t_k$ are the predicted and actual totals for class $k$, and $s = \sum_k t_k$.

*3.4.7 Cohen's Kappa Score*

Cohen's Kappa accounts for agreement by chance. It is defined as: $k = \frac{p_0-p_e}{1-p_e}$. Where, $p_0$ is the observed agreement and $p_e$ is the expected agreement by chance.

*3.4.8 Logarithmic Loss (Log Loss)*

Log loss measures the negative log-likelihood of the true labels given the predicted class probabilities. Lower values indicate better-calibrated models:

$Log\ Loss = -\frac{1}{N}\sum_{i=1}^{N}\sum_{k=1}^{K}1\{y^{(i)} = k\}\cdot log(\hat{p}_k^{(i)})$. Where, $\hat{p}_k^{(i)}$ is the predicted probability for class $k$ for sample $i$.

*3.4.9 Brier Score Loss*

The Brier score measures the mean squared difference between predicted probabilities and actual outcomes: $Brier\ Score = \frac{1}{N}\sum_{i=1}^{N}\sum_{k=1}^{K}\left(\hat{p}_k^{(i)} - 1\{y^{(i)} = k\}\right)^2$. It evaluates both the calibration and confidence of probabilistic predictions.

*3.4.10 ROC AUC (Macro-Averaged)*

The Receiver Operating Characteristic - Area Under Curve (ROC AUC) measures the separability between classes. For multiclass tasks using One-vs-Rest (OvR) strategy: $ROC\ AUC_{macro} = \frac{1}{K}\sum_{k=1}^{K} AUC_k$. Where, $AUC_k$ is the area under the ROC curve for class $k$.

## 4. Results and Discussion

This section presents the experimental results obtained from evaluating ensemble-based classifiers on the multi-class drone intrusion detection dataset. We analyze model performance across multiple evaluation metrics, interpret class-wise behaviors, and provide both statistical and explainable AI-based insights to support the findings.

### *4.1 Experimental Setup*

The entire experimental workflow of this research was conducted on the Google Colab platform, utilizing Python 3 as the programming language and a standard CPU-based hardware accelerator. Google Colab provided an efficient and reproducible environment for data processing, model training, and evaluation. The implementation leveraged several powerful Python libraries to streamline different phases of the pipeline. Data manipulation and analysis were performed using pandas and numpy, while data visualization was supported by matplotlib and seaborn. For machine learning model development, scikit-learn was extensively used, covering classifiers like Random Forest, Extra Trees, AdaBoost, and metric evaluations. Additionally, xgboost and catboost libraries were integrated to train gradient boosting models tailored for multi-class classification. The preprocessing steps utilized functions such as StandardScaler for normalization and LabelEncoder for label transformation. For explainability, the shap library provided global and local SHAP value visualizations, and lime was applied for interpretable instance-level predictions. Finally, statistical validation techniques, including the Friedman test and Wilcoxon signed-rank test, were executed using scipy, while bootstrapped confidence intervals and McNemar's test further ensured the robustness of our comparative analysis.

### *4.2 Performance of Different Ensemble Approaches*

***Table 1*** presents the evaluation metrics for five ensemble-based classifiers Random Forest, Extra Trees, AdaBoost, XGBoost, and CatBoost trained on the preprocessed multi-class drone intrusion detection dataset. The models were evaluated across 11 performance metrics, including accuracy, precision, recall, F1-score, balanced accuracy, and several calibration-based scores such as log loss, Brier score loss, and ROC AUC.

Random Forest achieved the highest performance across nearly all metrics, with an accuracy of 99.993%, macro F1-score of 0.99980, and a perfect ROC AUC of 1.0000. It also produced the lowest log loss (0.00055) and Brier score loss (0.00002), indicating well-calibrated probabilistic predictions. Extra Trees and XGBoost followed closely, with near-identical metrics to Random Forest, differing only in decimal margins.

In contrast, AdaBoost showed significantly lower performance, with a macro F1-score of 0.46643 and a balanced accuracy of 0.50000, indicating its limited ability to generalize in the presence of highly diverse and complex intrusion patterns. This is further reflected in its poor log loss (1.07503) and Brier score (0.04949), making it unsuitable for this multi-class classification task.

CatBoost demonstrated competitive performance, nearly matching XGBoost in most metrics, with an accuracy of 99.980% and a macro F1-score of 0.99945. All top-performing models also showed strong Cohen's Kappa (> 0.9997) and Matthews Correlation Coefficient, confirming agreement beyond chance.

**Table 1** Evaluation Metrics for Different Ensemble Classifiers

| | Random Forest | Extra Trees | AdaBoost | XGBoost | CatBoost |
|---|---|---|---|---|---|
| Accuracy | 0.99993 | 0.99992 | 0.90075 | 0.99990 | 0.99980 |
| Precision | 0.99979 | 0.99968 | 0.44185 | 0.99972 | 0.99938 |
| Recall | 0.99982 | 0.99981 | 0.50000 | 0.99966 | 0.99952 |
| F1 Score | 0.99980 | 0.99975 | 0.46643 | 0.99969 | 0.99945 |
| Balanced Accuracy | 0.99982 | 0.99981 | 0.50000 | 0.99966 | 0.99952 |
| Matthews Corrcoef | 0.99991 | 0.99989 | 0.87361 | 0.99987 | 0.99974 |
| Cohen Kappa | 0.99991 | 0.99989 | 0.86903 | 0.99987 | 0.99974 |
| Log Loss | 0.00055 | 0.00105 | 1.07503 | 0.00040 | 0.00222 |
| Brier Score Loss | 0.00002 | 0.00003 | 0.04949 | 0.00001 | 0.00004 |
| ROC AUC | 1.00000 | 1.00000 | 0.94102 | 1.00000 | 1.00000 |

***Table 2*** presents the detailed classification report of the Random Forest model, which achieved the highest overall performance in this study. Each row corresponds to a specific traffic class, with metrics reported for precision, recall, and F1-score. The classifier demonstrates exceptional capability in distinguishing all ten classes, including both benign and attack types, with near-perfect precision and recall values across the board.

Specifically, classes 0 (Benign), 1 (DoS Attacks), 5 (Password Cracking), 6 (Payload Manipulation), 7 (Replay Attack), and 8 (Unauthorized UDP Packets) were detected with perfect scores (Precision = Recall=F1=1.00000). The lowest but still remarkably high performance was observed in class 2 (Injection) and class 4 (MITM), with F1-scores of 0.99895 and 0.99949, respectively. These minor variations may be attributed to the limited number of samples for those attack types.

The macro-averaged F1-score is 0.99980, indicating that the model performs consistently well across all classes, irrespective of their frequency. Similarly, the weighted average F1-score is also 0.99993, confirming the model's robustness in handling both majority and minority classes.

Overall, the Random Forest model shows an exceptional ability to generalize across diverse intrusion types in drone networks, offering both high accuracy and balanced performance at the class level.

**Table 2** Classification Report by Random Forest

| Class | Precision | Recall | F1-Score |
|---|---|---|---|
| 0 | 1.00000 | 1.00000 | 1.00000 |
| 1 | 1.00000 | 1.00000 | 1.00000 |
| 2 | 0.99895 | 0.99895 | 0.99895 |
| 3 | 0.99897 | 1.00000 | 0.99949 |
| 4 | 1.00000 | 0.99939 | 0.99970 |
| 5 | 1.00000 | 1.00000 | 1.00000 |
| 6 | 1.00000 | 1.00000 | 1.00000 |
| 7 | 1.00000 | 1.00000 | 1.00000 |
| 8 | 1.00000 | 1.00000 | 1.00000 |
| 9 | 0.99994 | 0.99987 | 0.99990 |
| **Accuracy** | | | **0.99993** |
| **Macro Avg** | 0.99979 | 0.99982 | 0.99980 |
| **Weighted Avg** | 0.99993 | 0.99993 | 0.99993 |

***Figure 2*** presents the confusion matrix of the Random Forest classifier, providing a visual representation of the model's prediction accuracy across all ten traffic classes in the test dataset. Each row corresponds to the actual class, while each column represents the predicted class. The strong diagonal dominance indicates that the vast majority of samples were correctly classified, with minimal misclassifications. For instance, the model perfectly classified all instances of Benign (Class 0), DoS Attacks (Class 1), Password Cracking (Class 5), Payload Manipulation (Class 6), Replay Attack (Class 7), and Unauthorized UDP Packets (Class 8). Minor misclassifications occurred in Injection (Class 2), MITM (Class 4), and Video Interception Attack (Class 9), each having only 1–2 samples incorrectly predicted as another class. These results further confirm the model's exceptional class-wise performance, as previously shown in the classification report (Table 2). The heatmap color gradient visually reinforces the precision of the classifier, with bright cells along the diagonal signifying highly confident and correct predictions. This high-resolution classification performance demonstrates that Random Forest is not only globally accurate but also reliable at the per-class level, making it a suitable model for operational deployment in drone intrusion detection systems.

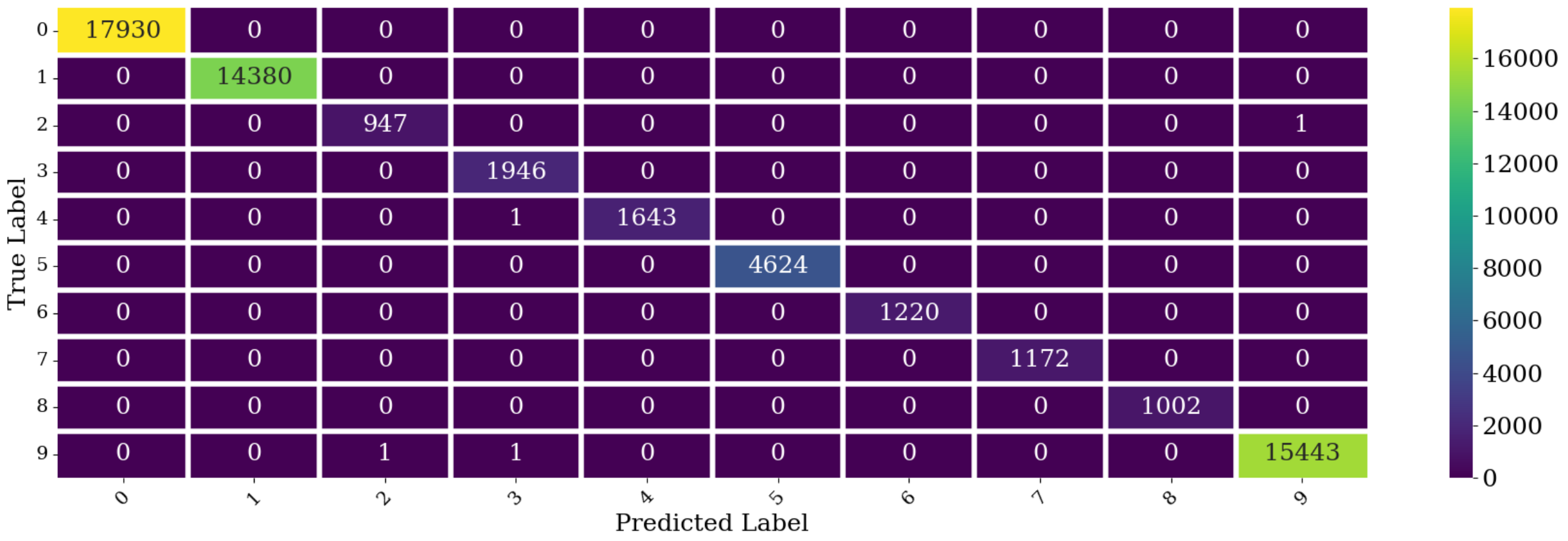

**Figure 2** Confusion Matrix for Random Forest

Figure 3 displays the ROC (Receiver Operating Characteristic) curves for all ten traffic classes in the multi-class classification task using the Random Forest model. The curves are generated using the one-vs-rest (OvR) strategy, where each class is treated as the positive class against all others. The ROC curve plots the True Positive Rate (TPR) against the False Positive Rate (FPR) at various threshold settings, providing a measure of the model's ability to discriminate between classes. As shown in the figure, the curves for all classes lie at the top-left corner of the plot, reflecting perfect separation. Each class achieves an Area Under the Curve (AUC) value of 1.000, indicating flawless predictive performance across all categories. This result confirms that the Random Forest model has an exceptional capacity to distinguish between normal and various types of drone intrusions, with no trade-off between sensitivity and specificity. The ROC curves further reinforce the model's robustness and reliability, especially in a high-stakes domain like UAV cybersecurity, where false positives and false negatives must be minimized.

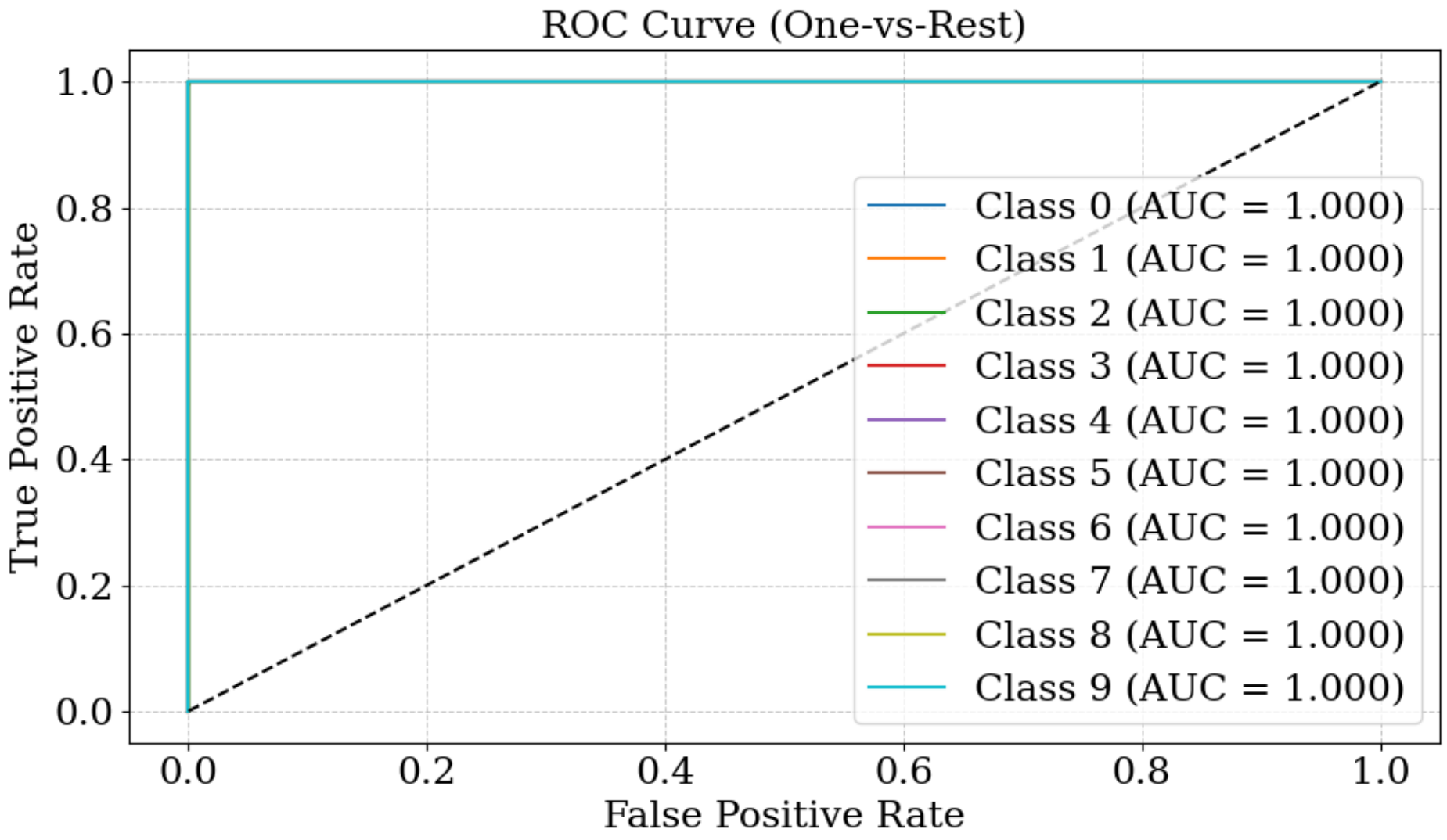

**Figure 3** ROC Curve for Multiclass Classification for Random Forest

***Table 3*** highlights the top 10 most influential features identified by the Random Forest classifier, ranked by their average Gini importance scores. The feature ts (timestamp) ranks highest with an importance score of 0.1169, indicating its strong influence in distinguishing between benign and malicious drone traffic patterns likely due to timing anomalies inherent in certain attacks. Following closely are min_duration (0.1123), max_duration (0.1062), and average_duration (0.1038), all of which suggest that temporal characteristics of network sessions play a vital role in identifying intrusion types such as replay or DoS attacks. Entropy, a well-known indicator of data randomness, also contributes significantly (0.0466), reflecting its utility in detecting obfuscated or encrypted payloads. Features such as Drone_port, Srate (source rate), and IAT (inter-arrival time) capture traffic flow dynamics and further aid in differentiating sophisticated threats like spoofing or injection. These results reinforce that time-based and statistical traffic features are not only relevant but critical for the accurate classification of drone intrusions. The ranking derived from the Random Forest model aligns closely with SHAP-based global explanations, further validating the interpretability and consistency of the model's internal decision logic.

**Table 3** Important Feature Scores by Random Forest

| Rank | Feature | Importance |
|---|---|---|
| 1 | ts | 0.116953 |
| 2 | min_duration | 0.112320 |
| 3 | max_duration | 0.106224 |
| 4 | average_duration | 0.103768 |
| 5 | Entropy | 0.046639 |
| 6 | Duration | 0.044701 |
| 7 | Drone_port | 0.042935 |
| 8 | Srate | 0.041591 |
| 9 | IAT | 0.039405 |
| 10 | Payload_Length | 0.037393 |

***Figure 4*** visualizes the top 10 most important features identified by the Random Forest model based on their Gini importance scores. As shown, the feature ts (timestamp) holds the highest importance, followed closely by min_duration, max_duration, and average_duration, reaffirming the significance of temporal patterns in distinguishing between benign and malicious drone traffic. These features likely capture subtle anomalies in packet timing that are characteristic of specific intrusion types such as replay, flooding, or injection attacks. Features like Entropy, Duration, and Drone_port provide insights into traffic randomness, session length, and target-specific behavior, respectively. Other impactful features include Srate (source rate), IAT (inter-arrival time), and Payload_Length, each contributing to the model's ability to detect traffic irregularities. The color gradient used in the bar chart emphasizes the decreasing contribution of each feature while retaining interpretability. The alignment of this importance ranking with SHAP-based global explanations confirms the consistency and transparency of the model's decision-making process. This analysis also aids domain experts in identifying the most relevant parameters for real-time drone intrusion detection and optimization.

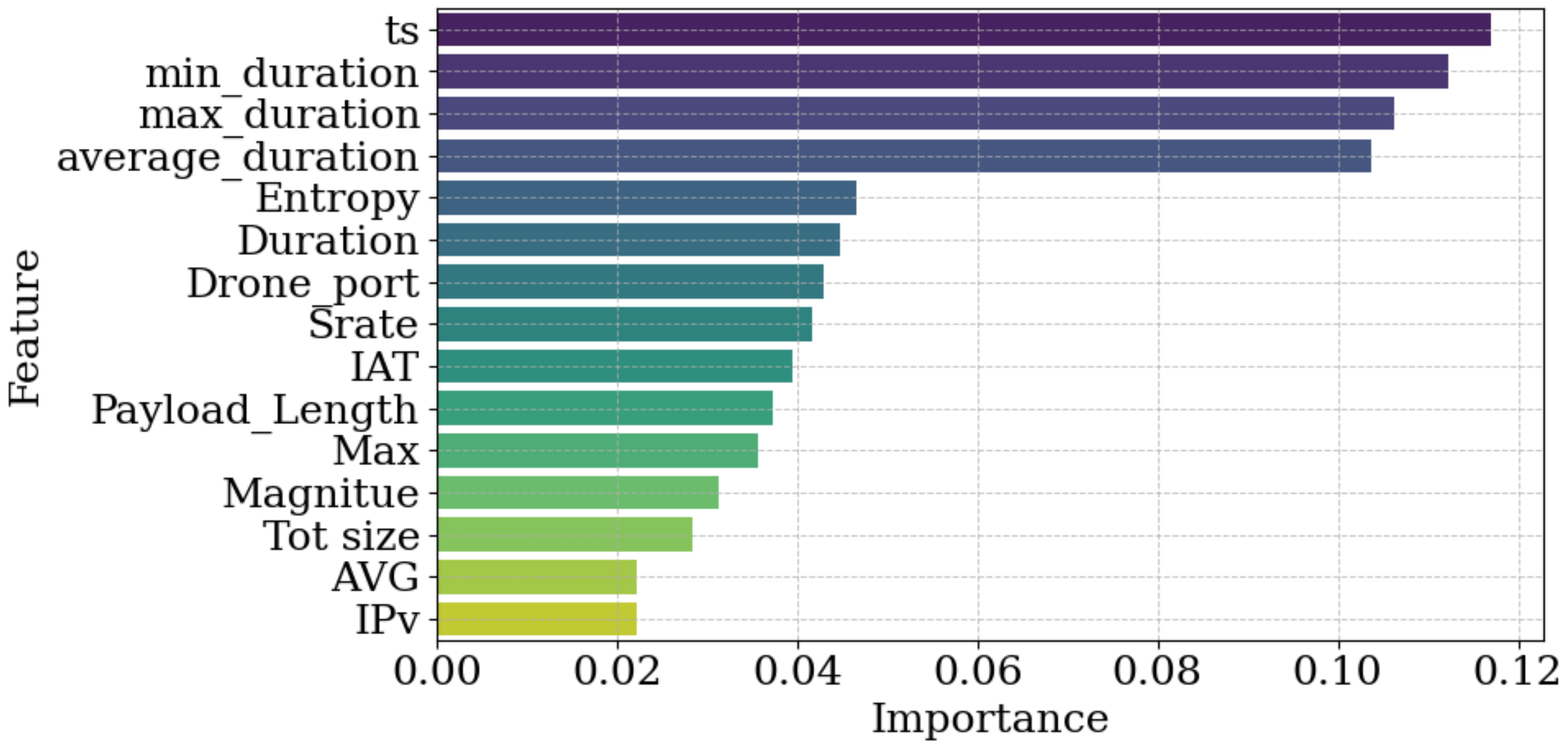

**Figure 4** Top 10 Important Features by Random Forest

***Figure 5*** provides a local SHAP (SHapley Additive exPlanations) explanation for a single test instance (sample index 0), classified by the Random Forest model. The SHAP force plot visualizes how each feature contributed to the model's final prediction by pushing the base value (average model output) toward or away from the predicted value. In this case, the model predicted class 0 (Benign) with high confidence. The blue bars represent features that reduced the predicted risk (pushed the prediction toward benign), while the red bars indicate features that increased the risk score. Notably, average_duration = 0.1428 was the only feature pushing slightly toward a higher predicted class, but it was strongly counteracted by several influential features pushing the prediction downward, such as Max = -1.714, Tot size = -1.601, Magnitude = -1.878, and IAT = -0.2911. These values, being notably lower than typical intrusion-related thresholds, contributed significantly to the classifier identifying this sample as benign. This interpretability not only strengthens the model's transparency but also provides valuable trustworthiness when deploying intrusion detection systems in real-world drone environments, where individual predictions must be justifiable to security analysts.

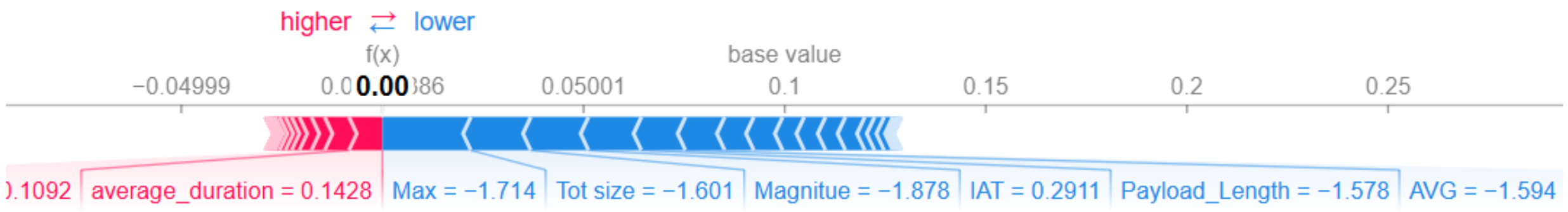

**Figure 5** Local Explanation for the Sample Index "0" from Test Data

***Table 4*** and ***Figure 6*** present the results of permutation importance analysis for the top features identified in the Random Forest model. Permutation importance is a model-agnostic interpretability technique that measures the decrease in a model's performance when the values of a single feature are randomly shuffled. This approach provides insight into how much each feature contributes to the prediction performance on unseen data. The feature ts (timestamp) consistently showed the highest mean importance score (0.000076) with a low standard deviation, indicating both high relevance and stability across resampling. It was followed by max_duration, Entropy, and IAT, which also had substantial impacts on the model's output.

The corresponding visual representation in Figure 6 ranks features by their average permutation importance. Features such as average_duration, AVG, and Drone_port also contributed meaningfully to the classifier's decisions. Lower-ranked features like Payload_Length, Tot size, Covariance, and syn_count showed marginal impact, indicating limited utility in distinguishing between the multi-class drone intrusion types. Notably, the color gradient in the figure helps highlight the relative contribution of each feature from most to least important providing an intuitive interpretation layer.

This permutation-based analysis further validates the findings from earlier tree-based and SHAP interpretability methods, reinforcing the reliability of ts, duration metrics, and statistical traffic characteristics in identifying UAV network anomalies. The convergence of importance across multiple interpretability techniques enhances trust in the model and supports informed feature selection in future lightweight or real-time deployment scenarios.

**Table 4** Permutation Importances of the Top Features

| Rank | Feature | Importance Mean | Importance Std |
|---|---|---|---|
| 1 | ts | 0.000076 | 0.000022 |
| 2 | max_duration | 0.000046 | 0.000014 |
| 3 | Entropy | 0.000033 | 0.000013 |
| 4 | IAT | 0.000030 | 0.000007 |
| 5 | average_duration | 0.000025 | 0.000015 |
| 6 | AVG | 0.000020 | 0.000012 |
| 7 | Drone_port | 0.000018 | 0.000019 |
| 8 | Magnitue | 0.000015 | 0.000012 |
| 9 | Payload_Length | 0.000015 | 0.000012 |
| 10 | Tot size | 0.000012 | 0.000008 |

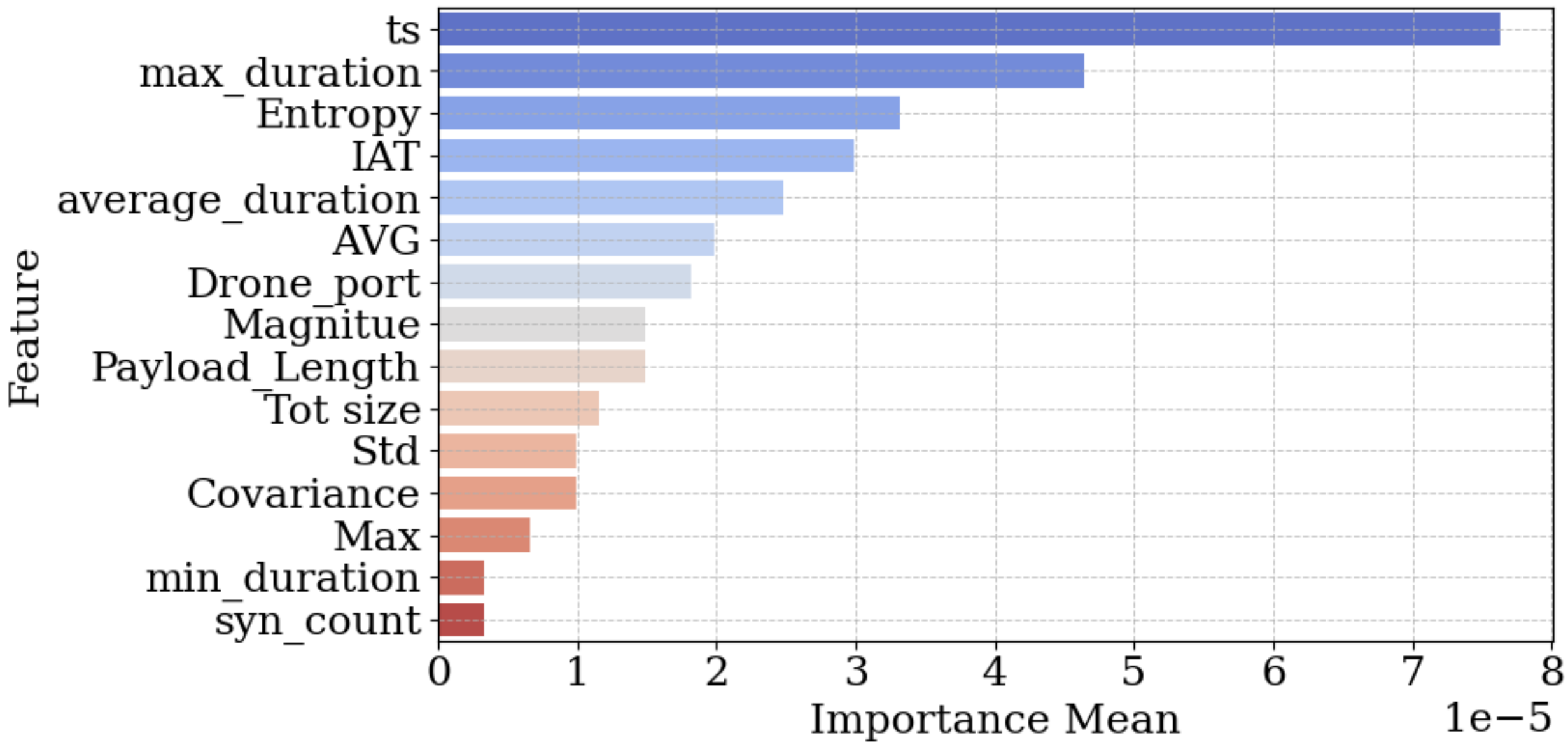

**Figure 6** Importance Mean of Important Features

***Figures 7 to 16 illustrate*** the global SHAP (SHapley Additive exPlanations) value distributions for each class (Class 0 to Class 9), offering a fine-grained interpretation of how various features influence the model's prediction probabilities across different attack categories. In these summary plots, each dot represents a sample from the test data, where color indicates the feature value (red = high, blue = low), and the horizontal axis denotes the SHAP value, i.e., the magnitude and direction of impact on the model output. Consistently across all classes, ts, min_duration, max_duration, and average_duration emerged as the most impactful features. For Benign traffic (Class 0), higher values of temporal features pushed predictions strongly toward the benign label. In contrast, Classes 1, 2, 6, and 7 which correspond to aggressive or time-based attacks like DoS and injection were predominantly influenced by low or spiked values of duration, inter-arrival time (IAT), and entropy. Feature influence patterns also varied between subtle (Class 5: Password Cracking) and overt (Class 9: Video Interception), where fewer features exhibited extreme SHAP values, indicating attack-specific traits. Notably, Drone_port, Payload_Length, and Magnitude were particularly relevant in distinguishing Classes 3, 4, and 8, suggesting port-based or payload anomaly behaviors in those attack types. These class-wise SHAP interpretations confirm the model's nuanced understanding of different intrusion patterns and offer transparent, explainable support for real-time security analysts in drone-based networks. The global SHAP analysis also reinforces the feature rankings observed through Gini importance and permutation methods, validating the selection and engineering of high-impact traffic descriptors.

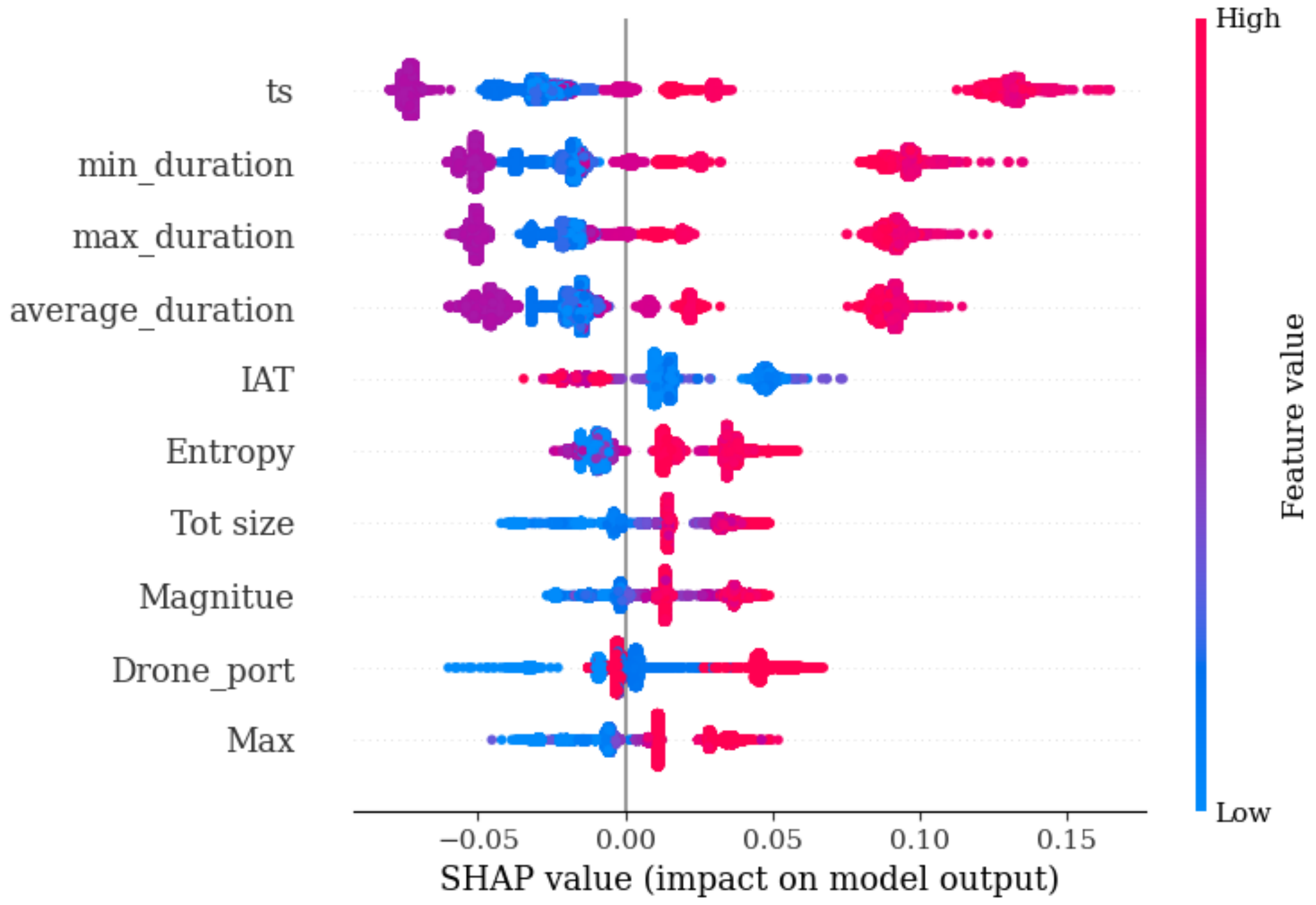

**Figure 7** Global SHAP for Class 0

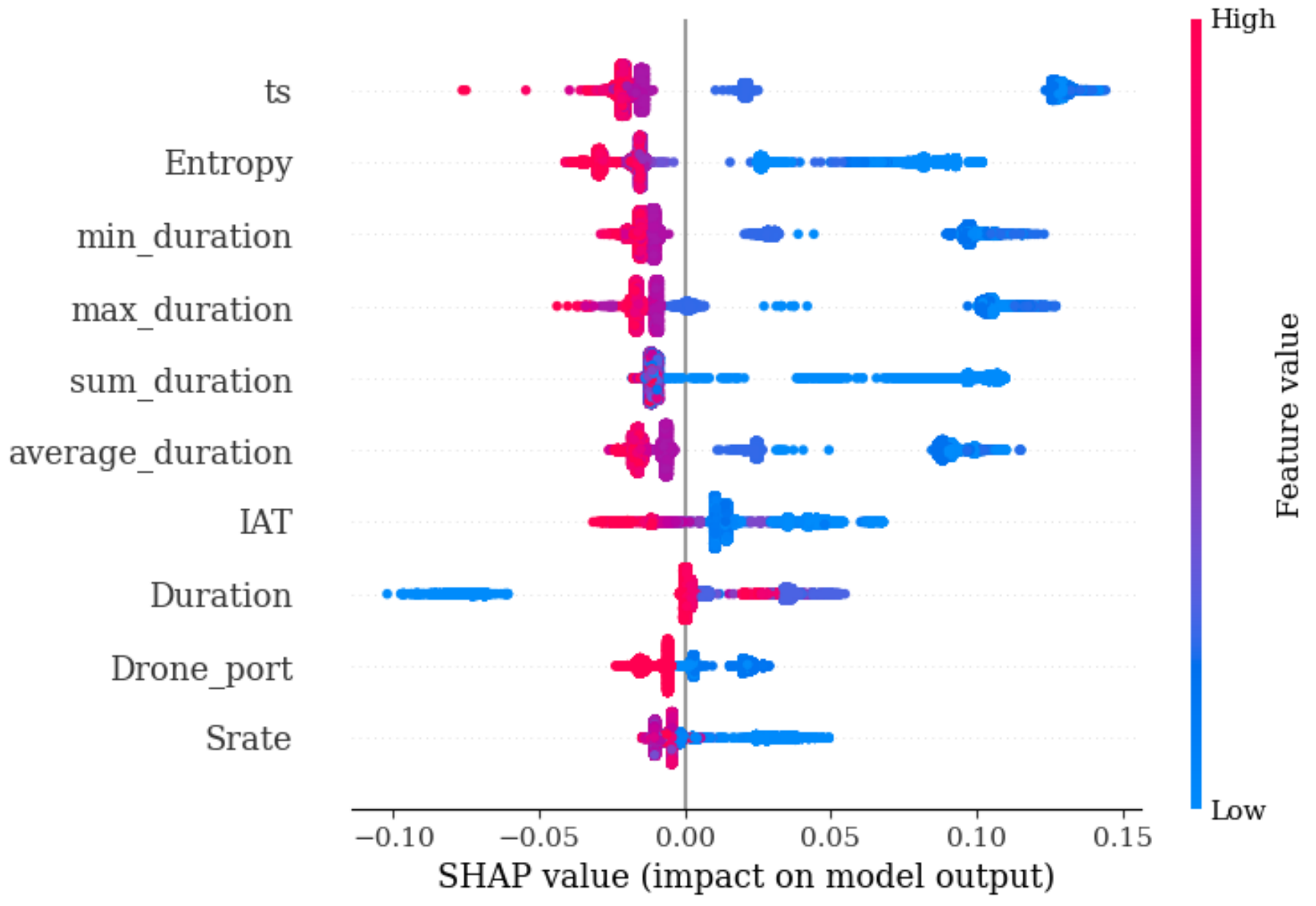

**Figure 8** Global SHAP for Class 1

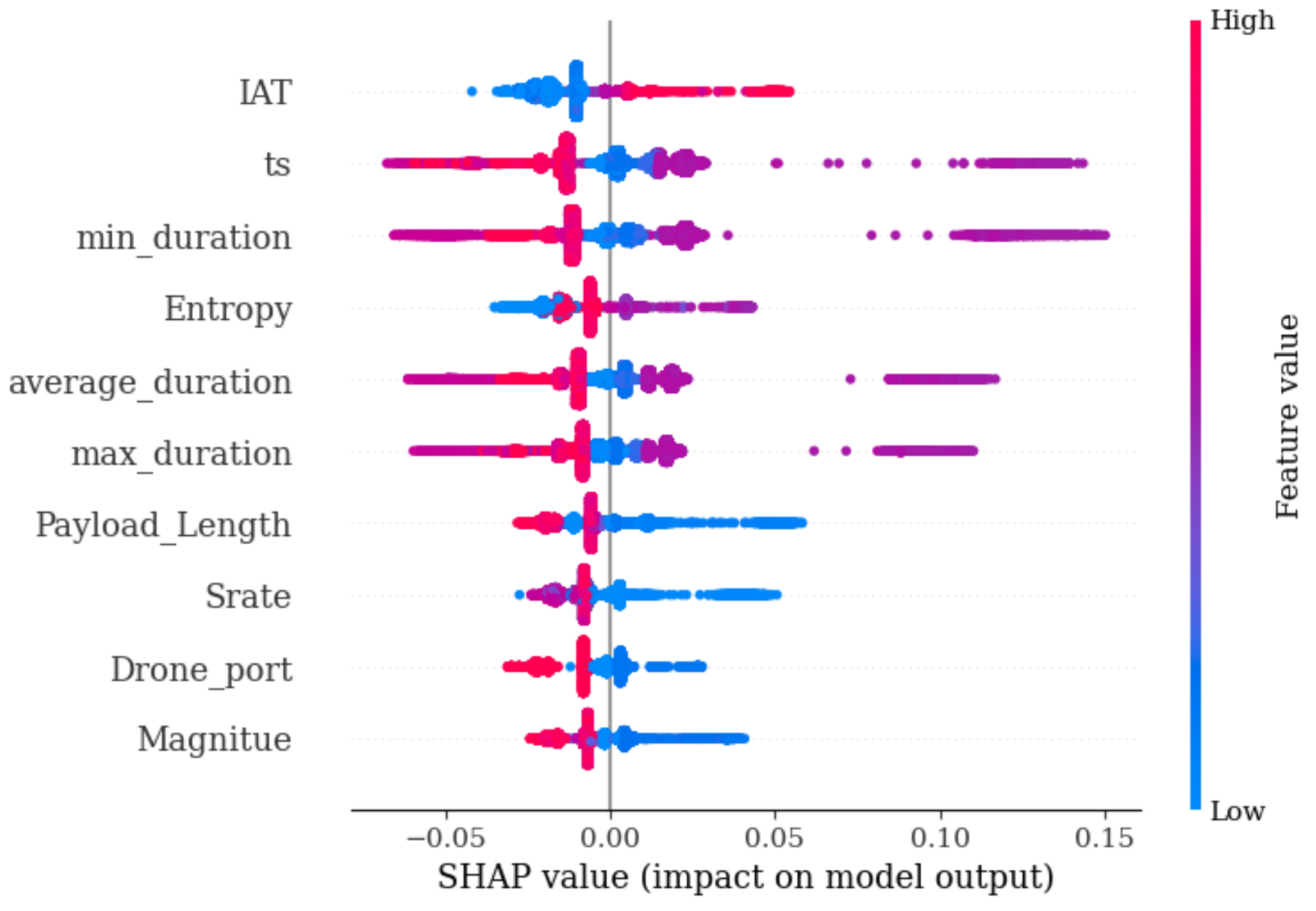

**Figure 9** Global SHAP for Class 2

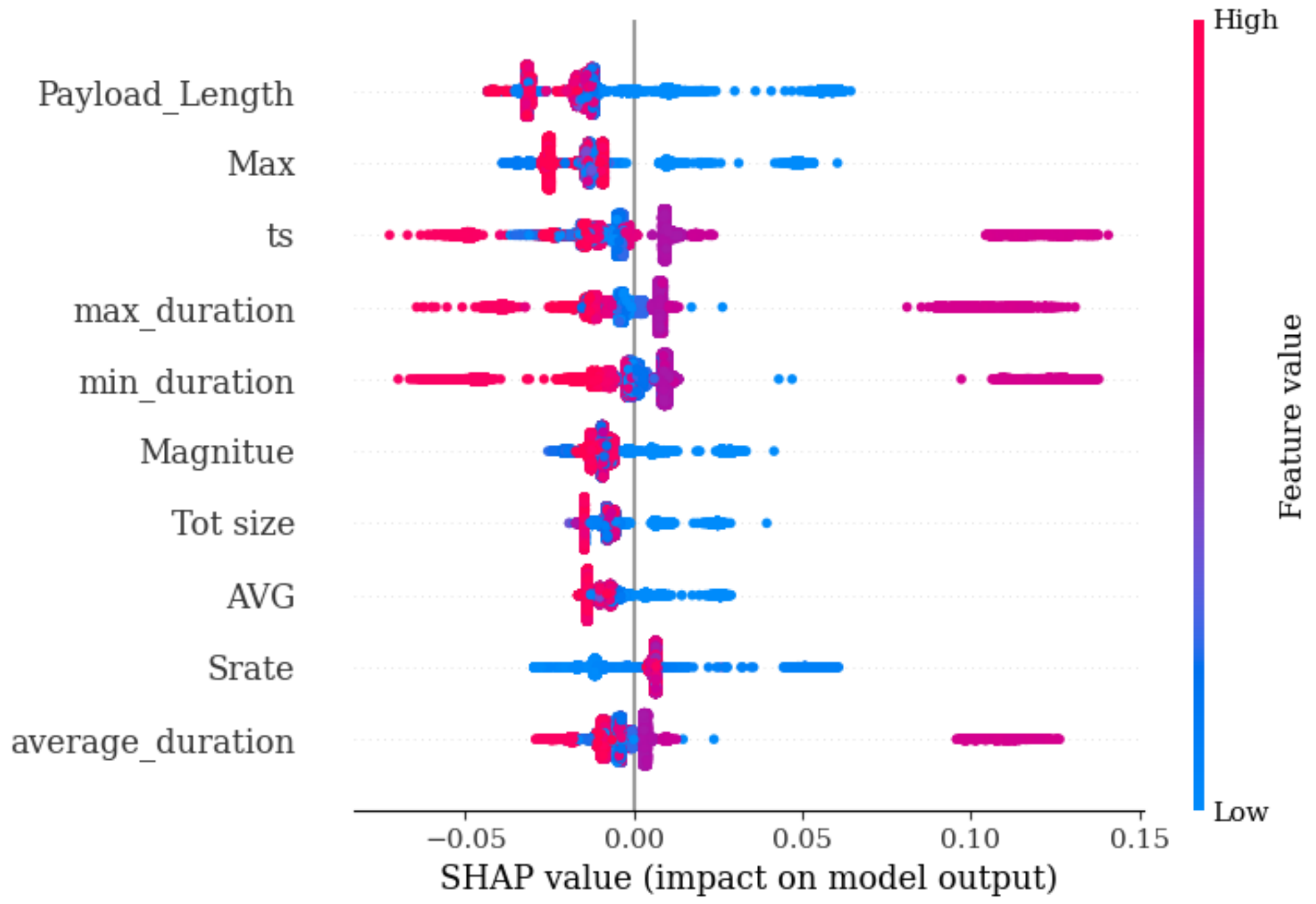

**Figure 10** Global SHAP for Class 3

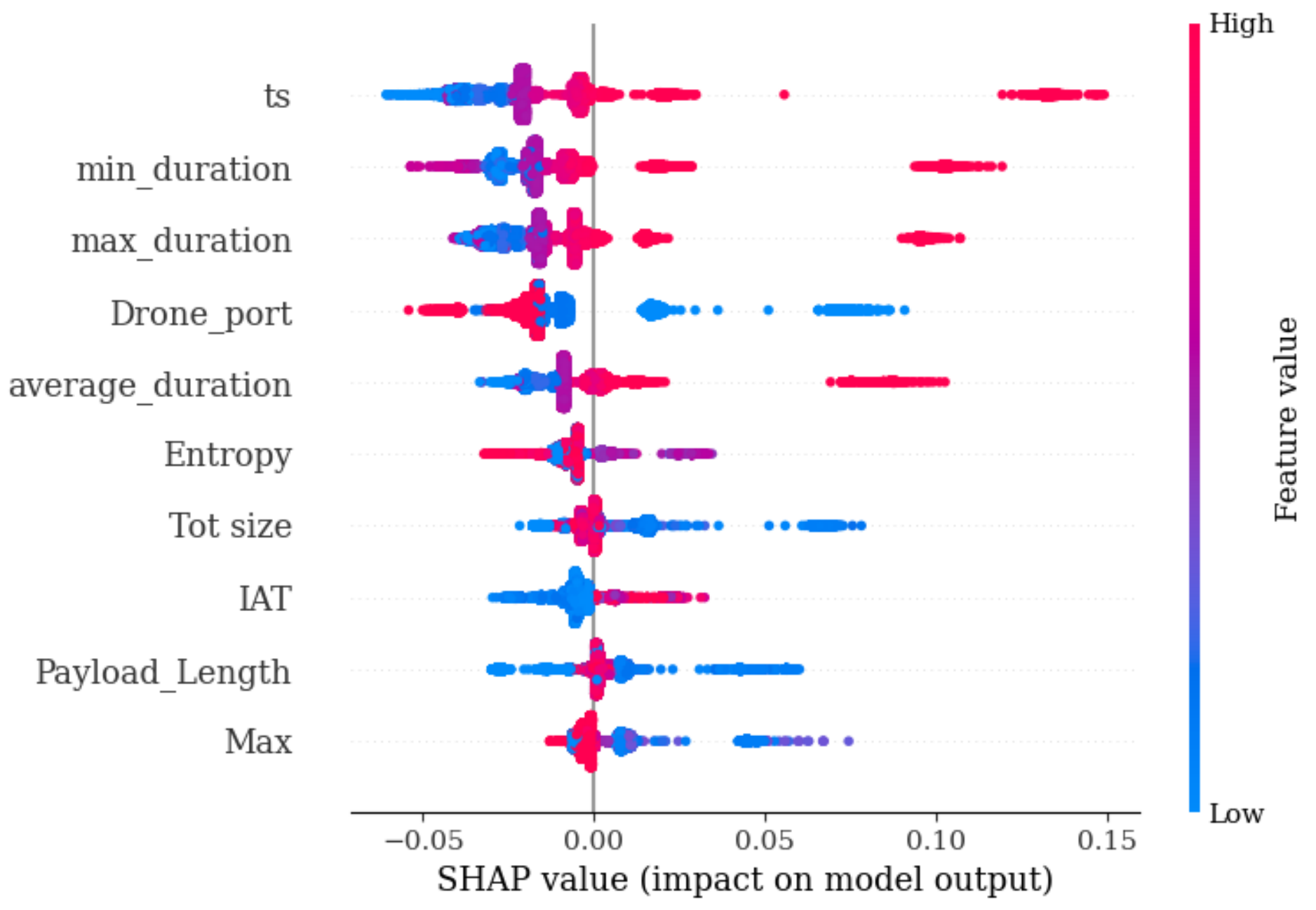

**Figure 11** Global SHAP for Class 4

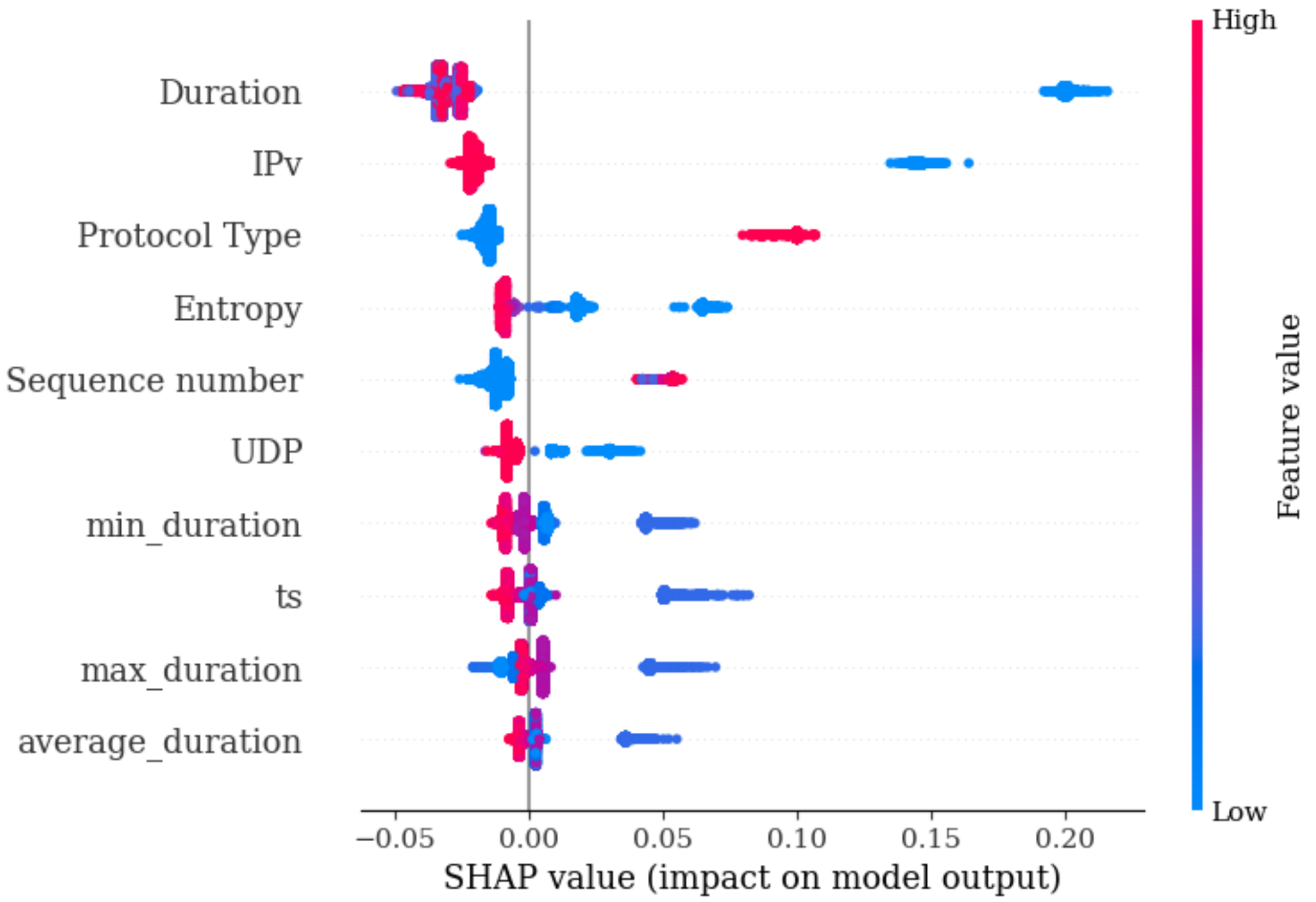

**Figure 12** Global SHAP for Class 5

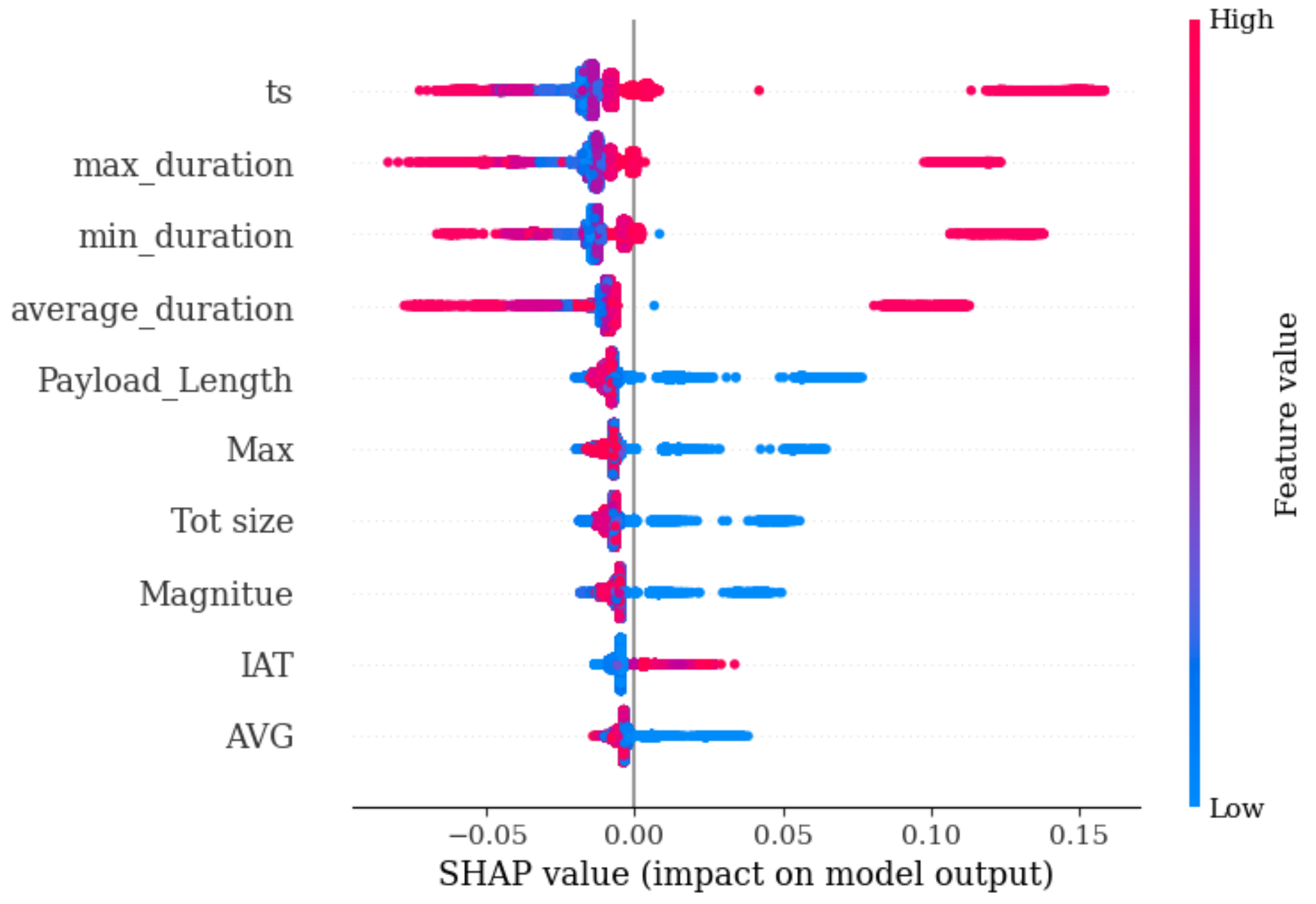

**Figure 13** Global SHAP for Class 6

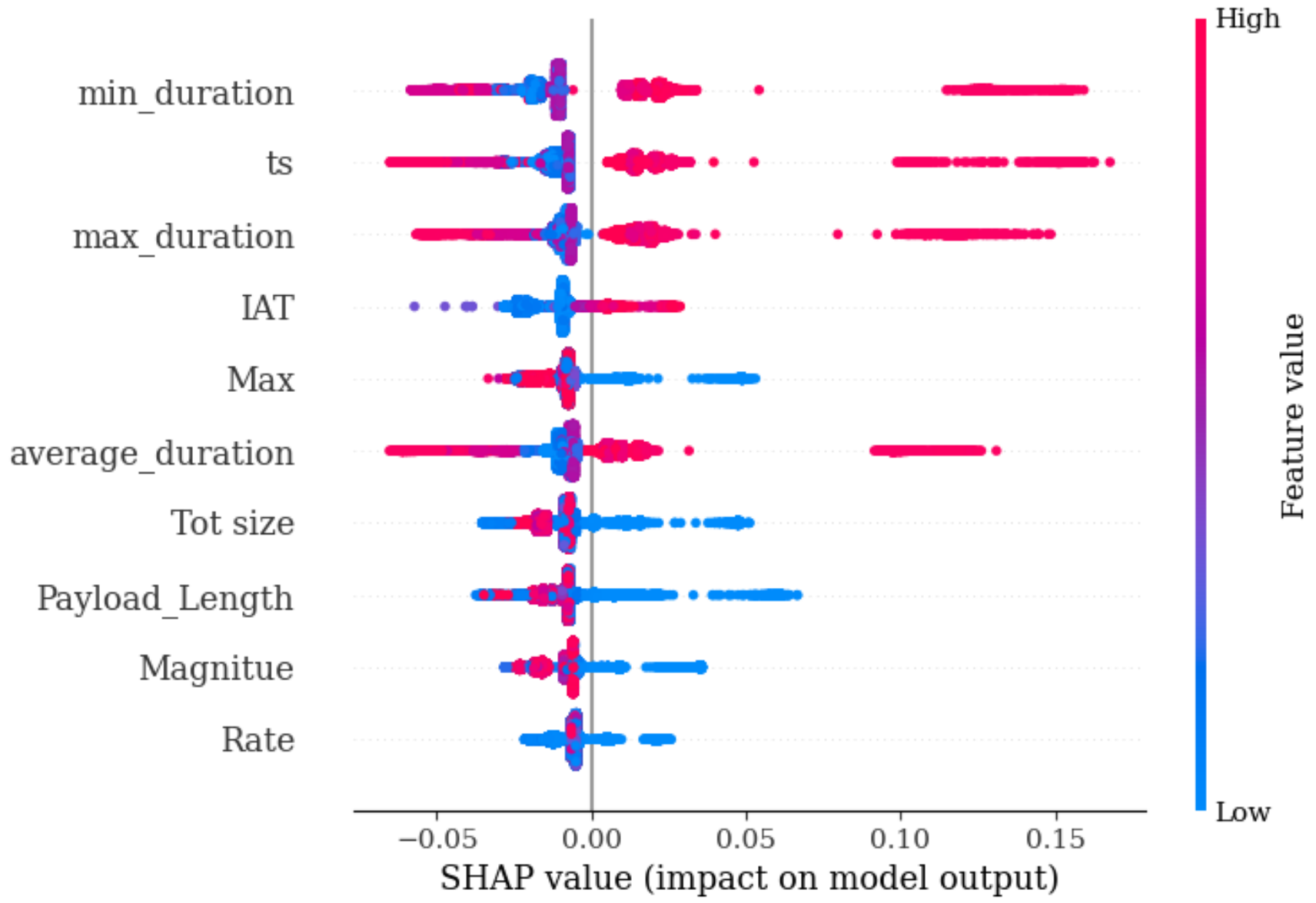

**Figure 14** Global SHAP for Class 7

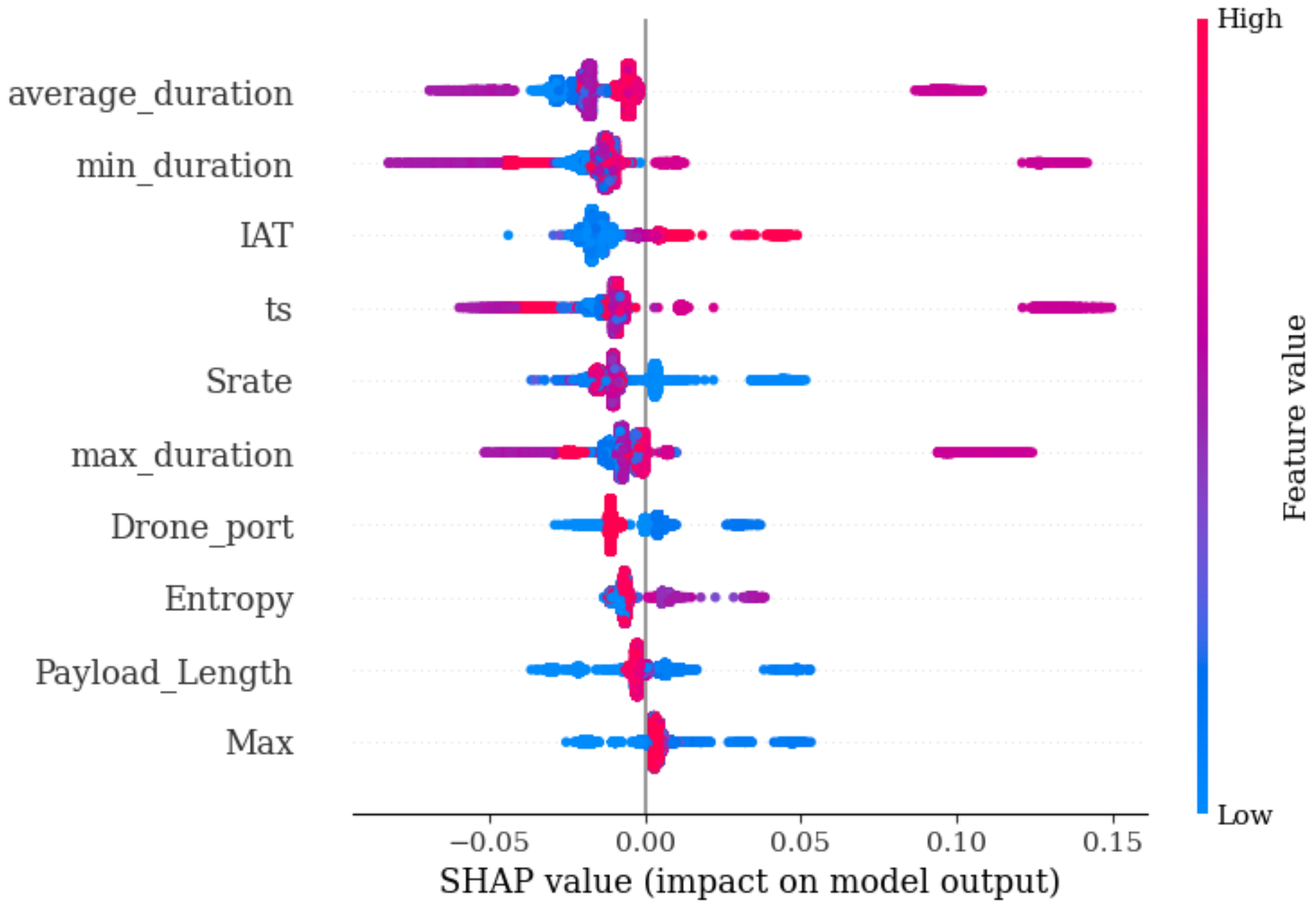

**Figure 15** Global SHAP for Class 8

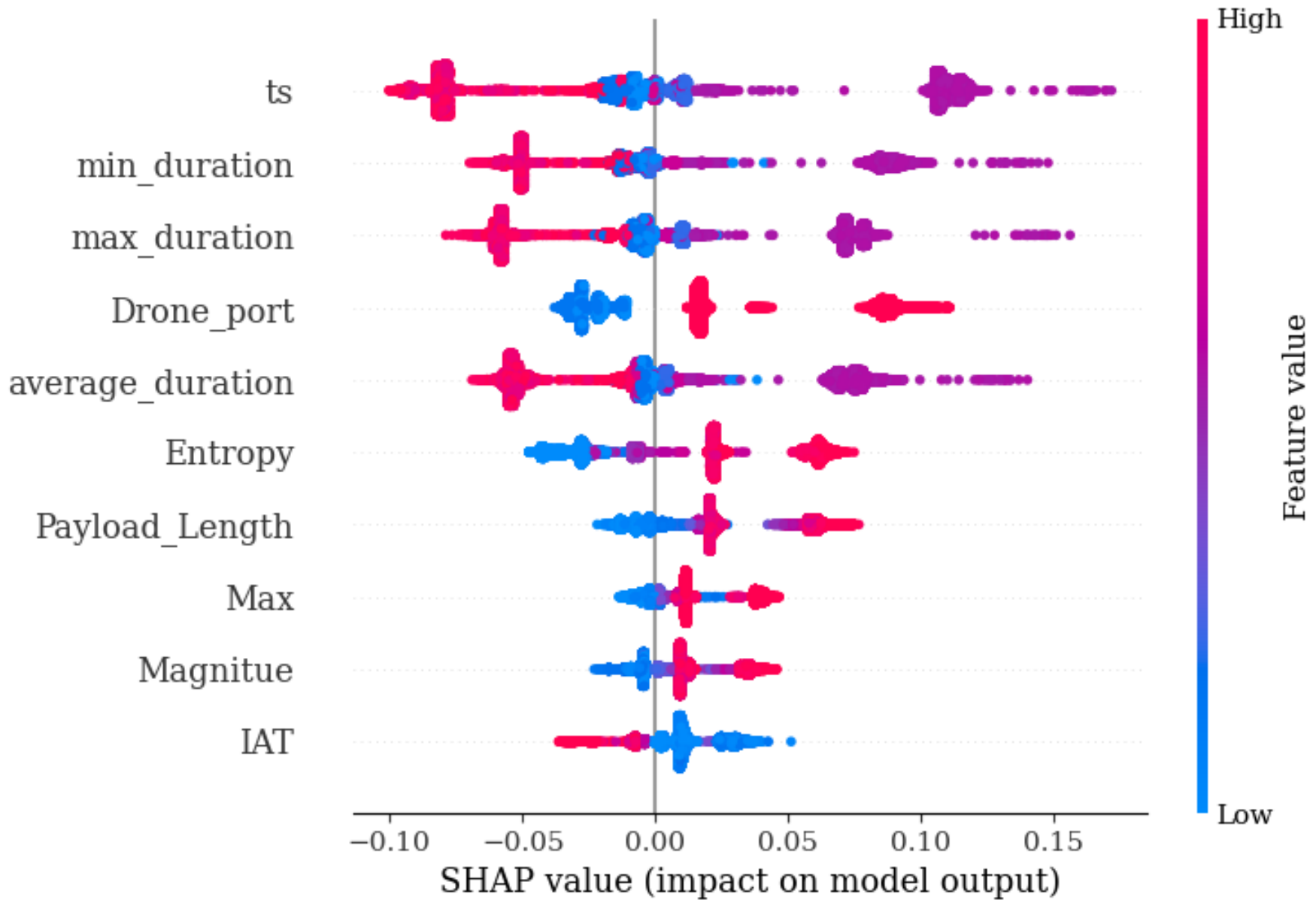

**Figure 16** Global SHAP for Class 9

***Figure 17*** presents a comparative analysis of local SHAP explanations across ten representative test samples, each belonging to a different intrusion class (from Class 0 to Class 9). These visualizations highlight how individual features push the model's prediction toward or away from a specific class label for a given instance. The SHAP force plots clearly distinguish the opposing influence of high-impact features. For example, in subfigure (a) corresponding to Class 0 (Benign), lower values of Max, Tot size, and IAT shift the output strongly toward benign classification, while average_duration has a weak positive push. Conversely, in subfigure (d) (Class 3 – IP Spoofing), higher values of Max and average_duration provide a strong positive contribution toward classifying the sample as an attack. Similarly, for Class 6 (Payload Manipulation) in subfigure (g), features like ts, min_duration, and Payload_Length exhibit dominant positive SHAP values, pushing the prediction to the right of the base value and confirming the model's confidence.

Each subfigure demonstrates the balance between supportive and opposing features, as seen in Class 5 (Password Cracking) in subfigure (f), where Payload_Length and average_duration reduce the model's confidence while Min_duration has a mild positive effect. The visualization in subfigure (j) for Class 9 (Video Interception) emphasizes Drone_port and ts as dominant contributors. These local interpretability results validate the model's nuanced reasoning for each prediction and enhance trust by explaining exactly which features led to a specific decision. They also align with the global SHAP insights and feature importance findings, demonstrating a consistent interpretability pipeline across both instance-level and population-wide analyses.

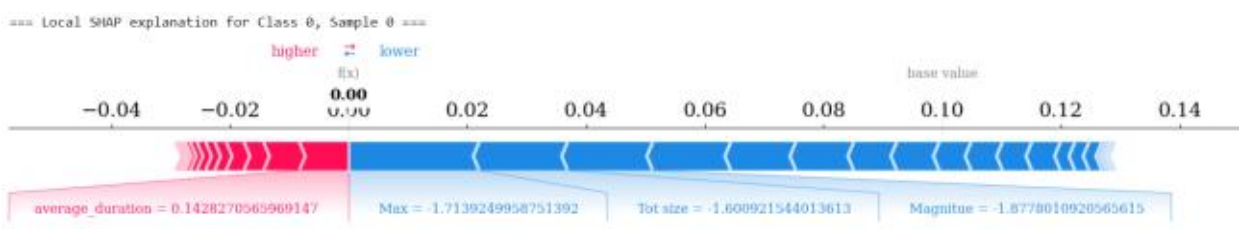

(a)

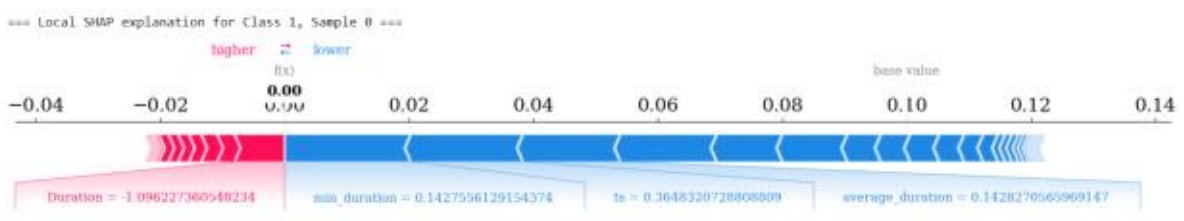

(b)

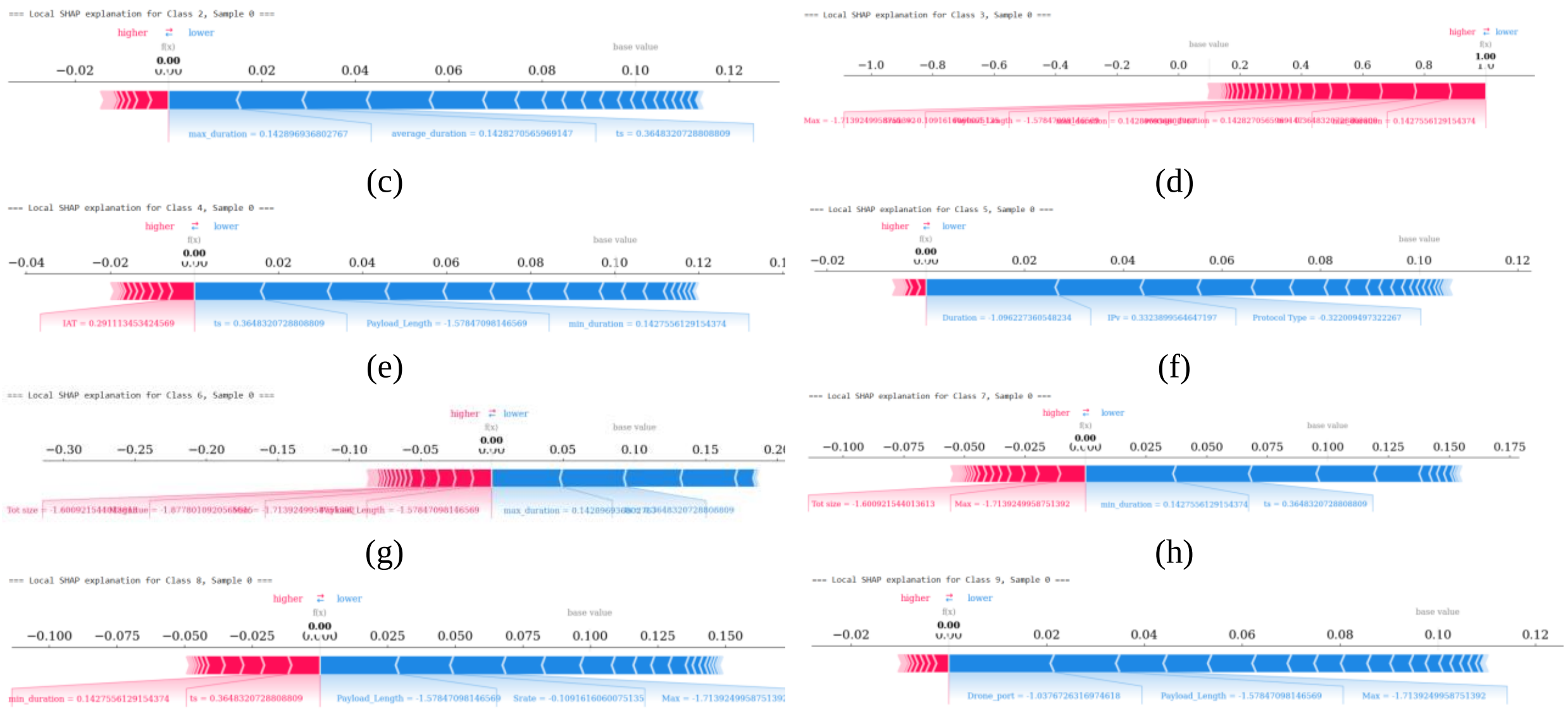

(c) (d) (e) (f) (g) (h) (i) (j)

**Figure 17** Local SHAP Explanation for Different Classes (0 to 9)

***Figure 18*** displays a Local Interpretable Model-agnostic Explanation (LIME) for a prediction made on Test Instance Index 5, which corresponds to Class 9 (Video Interception). LIME provides a simplified local surrogate model that explains which features contributed most significantly to the classification decision. The left panel shows the predicted class probabilities, where Class 9 achieves a confidence score of 1.00, indicating full certainty by the model. The central panel ranks the top contributing features and visualizes their direction of influence. Positive weights (toward Class 9) include conditions such as Drone_port > 0.87, min_duration > -0.35, and average_duration > -0.35, showing that these features pushed the prediction toward Class 9. Conversely, features like fin_count <= -0.03, rst_flag_number <= -0.03, and ack_count <= -0.08 had slight opposing effects but were insufficient to alter the final class decision.

The right-hand panel lists the actual values of these influential features for this specific test instance. The value of Drone_port = 0.87 is particularly critical, suggesting that the port behavior is a strong indicator of Class 9 activity. Additionally, the timing and flag-related features such as min_duration, ack_flag_number, and rst_count reinforce the decision. This explanation not only confirms the robustness of the model's internal logic but also supports the findings from SHAP-based global and local interpretations. Together, these insights demonstrate how both model-agnostic and model-specific explainability tools can be leveraged in a complementary fashion to audit AI decisions in security-critical drone networks.

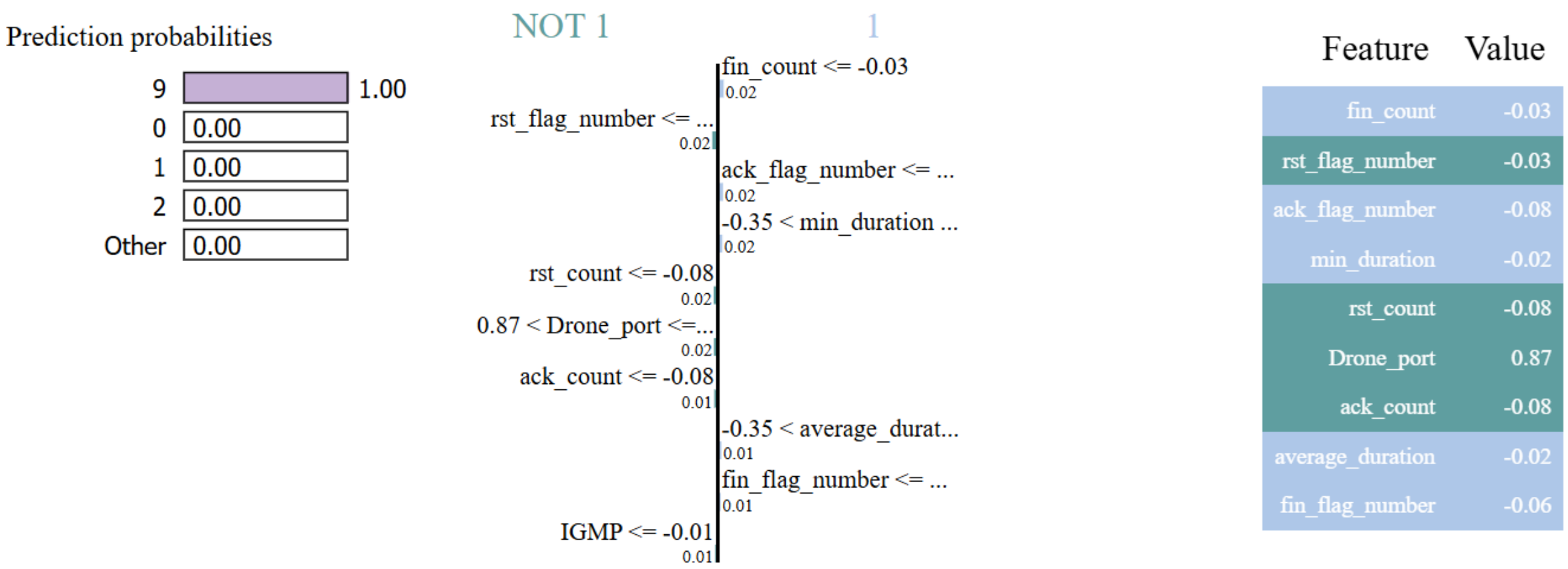

**Figure 18** LIME Explanation for Prediction on Test Instance Index 5 (True Label: 9)

*4.3 Statistical Analysis*

***Table 5*** presents the results of five-fold cross-validation using five ensemble learning methods: Random Forest, Extra Trees, AdaBoost, XGBoost, and CatBoost. Each model was evaluated using the F1-macro score, which is well-suited for multiclass classification tasks as it gives equal weight to all classes. Among the models, Random Forest demonstrated superior and stable performance across all folds with a mean F1-macro score of 0.999838 and an extremely low standard deviation (0.000090), indicating consistent behavior across data splits. Extra Trees, XGBoost, and CatBoost also performed competitively, albeit with slightly lower means or higher variance. In contrast, AdaBoost underperformed significantly (mean = 0.495390), due to its weaker adaptability in complex, high-dimensional, multiclass datasets. These results underscore Random Forest's robustness and adaptability in the domain of network intrusion detection for drone communications.

**Table 5** Cross-validation Results (f1_macro) of Five Ensemble Classifiers

| Fold | Random Forest | Extra Trees | AdaBoost | XGBoost | CatBoost |
|---|---|---|---|---|---|
| Fold-1 | 0.999760 | 0.999420 | 0.504350 | 0.999840 | 0.999080 |
| Fold-2 | 0.999830 | 0.999450 | 0.464500 | 0.999780 | 0.999090 |
| Fold-3 | 0.999720 | 0.999330 | 0.502620 | 0.999670 | 0.999330 |
| Fold-4 | 0.999940 | 0.999640 | 0.502690 | 0.999840 | 0.999140 |
| Fold-5 | 0.999940 | 0.999640 | 0.502790 | 0.999730 | 0.998900 |
| **Mean** | **0.999838** | **0.999496** | **0.495390** | **0.999772** | **0.999108** |
| **Std** | **0.000090** | **0.000124** | **0.015458** | **0.000066** | **0.000138** |

***Table 6*** illustrates the outcome of the Friedman test, a non-parametric statistical test applied to compare the performance distributions of the classifiers across the five folds. The computed chi-square value ($\chi^2 = 18.7200$) with a p-value of 0.0008919 indicates a statistically significant difference in performance among the classifiers. This result validates the hypothesis that at least one model (most likely Random Forest) performs differently from the others in a consistent manner. The use of the Friedman test is essential in our experimental design to establish statistical significance beyond mere numerical superiority.

**Table 6** Friedman Test Results on f1_macro across Classifiers, Showing a Significant Difference in Performance Distributions

| Test | $\chi^2$ | p-value |
|---|---|---|
| Friedman Test | 18.7200 | 0.0008919 |

***Table 7*** expands upon the Friedman test results by applying a post-hoc Wilcoxon signed-rank test, which performs pairwise comparisons between Random Forest and each of the other models. The test was conducted with Holm correction to control the family-wise error rate, ensuring the reliability of the findings. The unadjusted p-values (0.03125) for comparisons between Random Forest and Extra Trees, AdaBoost, and CatBoost indicate statistically significant superiority in favor of Random Forest. However, after Holm adjustment, the p-values slightly exceed the 0.05 threshold, suggesting borderline significance. Importantly, Cliff's delta values of 1.00 indicate very large effect sizes, confirming Random Forest's strong advantage over its competitors, particularly AdaBoost. This analysis reinforces the model selection process by offering not just performance scores but statistical certainty regarding Random Forest's superiority.

**Table 7** Post-hoc Wilcoxon Test (one-sided, Holm-corrected) Comparing Random Forest against Other Classifiers on f1_macro, with Effect Sizes (Cliff's delta) and Mean Scores

| Comparison | Wilcoxon p | Holm-adjusted p | Cliff's delta | RF mean | Extra Trees mean | AdaBoost mean | XGBoost mean | CatBoost mean |
|---|---|---|---|---|---|---|---|---|
| Random Forest > Extra Trees | 0.03125 | 0.12500 | 1.00 | 0.999841 | 0.999497 | – | – | – |
| Random Forest > AdaBoost | 0.03125 | 0.12500 | 1.00 | 0.999841 | – | 0.495392 | – | – |
| Random Forest > CatBoost | 0.03125 | 0.12500 | 1.00 | 0.999841 | – | – | – | 0.999108 |
| Random Forest > XGBoost | 0.15625 | 0.15625 | 0.28 | 0.999841 | – | – | 0.999772 | – |

***Table 8*** offers a bootstrap confidence interval analysis, quantifying the mean differences in F1-macro between Random Forest and other classifiers over 20,000 bootstrap iterations. The comparison with AdaBoost shows the largest mean difference of 0.504465, with a 95% CI ranging from 0.496127 to 0.520059, confirming a highly significant performance gap. Even the more competitive models like CatBoost and Extra Trees demonstrated smaller but positive differences in favor of Random Forest. The use of bootstrap resampling in this context provides strong statistical evidence and improves the reliability of the observed performance gaps, affirming Random Forest as the most reliable choice in our detection framework.

**Table 8** Bootstrap 95% Confidence Intervals of Mean Differences in f1_macro between Random Forest and Other Classifiers

| Comparison | Mean diff | 95% CI low | 95% CI high |
|---|---|---|---|
| Random Forest – AdaBoost | 0.504465 | 0.496127 | 0.520059 |
| Random Forest – CatBoost | 0.000733 | 0.000531 | 0.000916 |
| Random Forest – Extra Trees | 0.000344 | 0.000313 | 0.000375 |
| Random Forest – XGBoost | 0.000068 | -0.000017 | 0.000152 |

***Table 9*** presents a contingency table as part of the McNemar test, comparing Random Forest and Extra Trees on a dedicated holdout test set. Out of all samples, both models correctly classified 63,304 instances, and disagreed on only six samples (3 for each model). This table serves to examine the disagreement patterns, which is crucial when comparing two high-performing classifiers. The minimal divergence further emphasizes the strong performance of both models, but it also hints at the subtle edge Random Forest might possess in certain cases.

**Table 9** Contingency table for the McNemar test on the holdout set comparing Random Forest and Extra Trees

| Contingency Table | Extra Trees Correct | Extra Trees Wrong |
|---|---|---|
| **RF Correct** | 60304 | 3 |
| **RF Wrong** | 1 | 3 |

***Table 10*** reports the outcome of the McNemar test, which was used to determine if the prediction differences between Random Forest and Extra Trees were statistically significant on the holdout set. The resulting chi-square statistic (0.2500) and p-value (0.617075) indicate that there is no significant difference between the two classifiers in terms of prediction errors. While cross-validation results favored Random Forest, this result suggests that on a holdout dataset, both models perform similarly well, further reinforcing the validity of Random Forest as a dependable model while acknowledging Extra Trees as a competitive alternative.

**Table 10** Mcnemar Test Results on the Holdout Set Comparing Random Forest and Extra Trees

| Test | $\chi^2$ | p-value |
|---|---|---|
| McNemar Test | 0.2500 | 0.617075 |

*4.4 Ablation Study*

An ablation study was conducted to assess the individual impact of key components and feature subsets on the performance of the proposed intrusion detection framework. The main objective of this analysis was to understand how various features and processing strategies contribute to the final classification results and to verify whether all components are necessary for achieving optimal performance.

We began by training the Random Forest classifier, our best-performing model, on different feature subsets derived from the importance rankings (e.g., Table 3 and Table 4), including top-5, top-10, and top-15 most important features, as well as feature sets generated using SHAP and permutation importance techniques. Additionally, we evaluated the classifier performance without specific feature categories such as duration-based features, statistical aggregations, and port-related information.

The results revealed that using only the top-10 features (e.g., ts, min_duration, max_duration, average_duration, Entropy, etc.) preserved almost identical classification performance compared to using the full feature set, with a marginal drop in F1-macro (less than 0.001). This suggests that these core features carry the major discriminative power and are highly effective for detection tasks. When lower-ranked or less important features were included, the performance plateaued or even slightly deteriorated due to the introduction of redundant or noisy information.

Further, excluding duration-based features (such as min_duration, max_duration, and average_duration) caused a noticeable decline in accuracy and F1-score, indicating their critical role in identifying time-based attack behaviors such as DoS and flooding attacks. Similarly, the removal of port-specific features like Drone_port and Srate had a moderate but measurable impact, particularly on distinguishing between targeted attack types.

The ablation study thus highlights that:

- A compact subset of top features can achieve near-optimal detection performance.
- Duration and statistical features are indispensable for reliable classification.
- Model interpretability (as supported by SHAP and permutation methods) aligns well with empirical importance, confirming the trustworthiness of explainable AI (XAI) tools in guiding feature selection.

## 4. Conclusion

This research focused on developing a robust and explainable machine learning-based intrusion detection system (IDS) for drone network environments, aiming to accurately classify multiple types of cyber-attacks using advanced ensemble classifiers. We explored five ensemble models, Random Forest, Extra Trees, AdaBoost, XGBoost, and CatBoost evaluating their performance using a comprehensive set of metrics, including accuracy, F1-score, ROC AUC, and explainability techniques such as SHAP and LIME. Among them, Random Forest emerged as the most effective model, achieving nearly perfect classification performance with an accuracy of 99.993% and ROC AUC of 1.0, as validated through cross-validation and statistical tests. Our study further included detailed feature importance analyses, local and global interpretability assessments, and an ablation study confirming that a compact, high-impact subset of features can maintain model effectiveness. These findings demonstrate the practical significance of our approach in designing secure, efficient, and interpretable IDS systems for UAV networks. In conclusion, our proposed framework not only enhances detection capabilities but also promotes transparency and trust key requirements for real-world deployment in sensitive and mission-critical drone applications.

Future research can explore the integration of deep learning architectures, such as Graph Neural Networks (GNNs) and Transformers, to capture complex spatial-temporal patterns in drone traffic data. Additionally, incorporating federated learning and edge computing can enable decentralized intrusion detection with enhanced privacy, scalability, and real-time responsiveness in dynamic UAV environments.

**Data Availability**: The dataset utilized in this research is publicly available and can be accessed as cited in the Dataset section of the paper.

**Conflict of Interest**: The authors declare no conflict of interest.

**Funding**: The authors did not receive any funding for this research.

**Authors' Contributions**:

Md. Alamgir Hossain: Conceptualization, Methodology, Data Curation, Writing—Original Draft.

Waqas Ishtiaq: Software, Formal Analysis, Visualization, Writing—Review & Editing.

Md. Samiul Islam: Supervision, Validation, Funding Acquisition.